\documentclass{ws-ijmpa}
\usepackage[super,compress]{cite}
\usepackage{graphicx}
\usepackage{hyperref}
\usepackage{mathrsfs}  
\usepackage{mathabx}
\usepackage[shortlabels]{enumitem}
\usepackage[utf8]{inputenc}

\begin{document}
\markboth{Daniel S. Park}{Recent developments in
2d $\mathcal{N}=(2,2)$ supersymmetric gauge theories}

\newcommand{\cA}{\mathcal{A}}
\newcommand{\cB}{\mathcal{B}}
\newcommand{\cC}{\mathcal{C}}
\newcommand{\cD}{\mathcal{D}}
\newcommand{\cE}{\mathcal{E}}
\newcommand{\cF}{\mathcal{F}}
\newcommand{\cG}{\mathcal{G}}
\newcommand{\cH}{\mathcal{H}}
\newcommand{\cI}{\mathcal{I}}
\newcommand{\cJ}{\mathcal{J}}
\newcommand{\cK}{\mathcal{K}}
\newcommand{\cL}{\mathcal{L}}
\newcommand{\cM}{\mathcal{M}}
\newcommand{\cN}{\mathcal{N}}
\newcommand{\cO}{\mathcal{O}}
\newcommand{\cP}{\mathcal{P}}
\newcommand{\cQ}{\mathcal{Q}}
\newcommand{\cR}{\mathcal{R}}
\newcommand{\calS}{\mathcal{S}}
\newcommand{\cT}{\mathcal{T}}
\newcommand{\cU}{\mathcal{U}}
\newcommand{\cV}{\mathcal{V}}
\newcommand{\cW}{\mathcal{W}}
\newcommand{\cX}{\mathcal{X}}
\newcommand{\cY}{\mathcal{Y}}
\newcommand{\cZ}{\mathcal{Z}}
\newcommand{\bA}{\mathbb{A}}
\newcommand{\bB}{\mathbb{B}}
\newcommand{\bC}{\mathbb{C}}
\newcommand{\bD}{\mathbb{D}}
\newcommand{\bE}{\mathbb{E}}
\newcommand{\bF}{\mathbb{F}}
\newcommand{\bG}{\mathbb{G}}
\newcommand{\bH}{\mathbb{H}}
\newcommand{\bI}{\mathbb{I}}
\newcommand{\bJ}{\mathbb{J}}
\newcommand{\bK}{\mathbb{K}}
\newcommand{\bL}{\mathbb{L}}
\newcommand{\bM}{\mathbb{M}}
\newcommand{\bN}{\mathbb{N}}
\newcommand{\bO}{\mathbb{O}}
\newcommand{\bP}{\mathbb{P}}
\newcommand{\bQ}{\mathbb{Q}}
\newcommand{\bR}{\mathbb{R}}
\newcommand{\bS}{\mathbb{S}}
\newcommand{\bT}{\mathbb{T}}
\newcommand{\bU}{\mathbb{U}}
\newcommand{\bV}{\mathbb{V}}
\newcommand{\bW}{\mathbb{W}}
\newcommand{\bX}{\mathbb{X}}
\newcommand{\bY}{\mathbb{Y}}
\newcommand{\bZ}{\mathbb{Z}}
\newcommand{\fA}{\mathfrak{A}}
\newcommand{\fB}{\mathfrak{B}}
\newcommand{\fR}{\mathfrak{R}}
\newcommand{\fg}{\mathfrak{g}}
\newcommand{\fm}{\mathfrak{m}}
\newcommand{\fn}{\mathfrak{n}}
\newcommand{\fS}{\mathfrak{S}}
\newcommand{\fu}{\mathfrak{u}}
\newcommand{\bfa}{\mathbf{a}}
\newcommand{\bfh}{\mathbf{h}}
\newcommand{\bfJ}{\mathbf{J}}
\newcommand{\bfQ}{\mathbf{Q}}
\newcommand{\bfT}{\mathbf{T}}
\newcommand{\bfy}{\mathbf{y}}
\newcommand{\bfx}{\mathbf{x}}

\newcommand{\tPhi}{\widetilde{\Phi}}
\newcommand{\fh}{\mathfrak{h}}
\newcommand{\fc}{\mathfrak{c}}
\newcommand{\bfG}{\mathbf{G}}
\newcommand{\bfGa}{\mathbf{\Gamma}}
\newcommand{\bfH}{\mathbf{H}}
\newcommand{\bfC}{\mathbf{C}}
\newcommand{\rk}{\text{rk}}
\newcommand{\ov}{\over}
\newcommand\acomm[2]{\left\{#1,#2 \right\}}

\def\ts{{\widetilde{\sigma}}}
\def\tl{\widetilde{\lambda}}
\def\tphi{\widetilde{\phi}}
\def\tpsi{\widetilde{\psi}}
\def\tF{\widetilde{F}}
\def\tH{\widetilde{\mathcal{H}}}
\def\tC{\widetilde{C}}

\def\tm{\widetilde{m}}
\def\tfm{\widetilde{\mathfrak{m}}}
\def\tW{\widetilde{W}}
\def\twr{\widetilde{r}}
\def\tes{\widetilde{s}}
\def\thm{\widetilde{\widehat{m}}}
\def\tp{\widetilde{\partial}}
\def\tom{\widetilde{\om}}
\def\teta{\widetilde{\eta}}
\def\tG{\widetilde{G}}

\def\hm{\widehat{m}}
\def\hr{\widehat{r}}
\def\hs{\widehat{s}}
\def\hS{\widehat{S}}
\def\hsig{\widehat{\sigma}}
\def\hSig{\widehat{\Sigma}}
\def\hW{\widehat{W}}

\def\thW{\widetilde{\widehat{W}}}
\def\tfT{\widetilde{\mathfrak{T}}}
\def\tfM{\widetilde{\mathfrak{M}}}

\def\cs{\widecheck{s}}
\def\cS{\widecheck{\Sigma}}
\def\rc{\widecheck{r}}
\def\cu{\widecheck{u}}
\def\cx{\widecheck{x}}

\def\bo{{\bar{1}}}

\def\Gr{\text{Gr}}
\def\UV{\text{UV}}
\def\eff{\text{eff}}
\def\re{\mathrm{Re}}
\def\tr{\mathrm{tr}}
\def\epsdef{{\mathbf{\epsilon}_\Omega}}

\def\sing{\text{sing}}
\def\evon#1{\big{|}_{#1}}
\def\ha{\frac{1}{2}}
\def\rR{\text{R}}
\def\rI{\text{I}}
\def\rN{\text{N}}
\def\rS{\text{S}}


\newcommand{\unit}{\mathbbm{1}}

\newcommand{\ie}{{i.e.}}
\newcommand{\eg}{{e.g.}}

\newcommand{\ex}{{\bf x}}
\newcommand{\why}{{\bf y}}
\newcommand{\eN}{{\bf N}}
\newcommand{\zee}{{\bf z}}
\newcommand{\Qyoo}{{\bf Q}}

\def\lam{{\lambda}}
\def\Lam{{\Lambda}}
\def\al{{\kappa}}
\def\ket#1{|#1\rangle}
\def\bra#1{\langle#1|}
\def\vev#1{\langle#1\rangle}
\def\Res#1{\mathop{\text{Res}}_{#1}}
\def\JKRes#1{\mathop{\text{JK-Res}}_{#1}}

\def\eq#1{(\ref{#1})}
\def\p{\partial}

\def\Om{{\Omega}}
\def \th{{\theta}}
\def\a{{\alpha}}
\def\b{{\beta}}
\def \lam {\lambda}
\def \om {\omega}
\def \ra {\rightarrow}
\def\sig{{\sigma}}
\def\ep{{\epsilon}}
\def\up{{\upsilon}}
\def\apr{{\kappa'}}
\def\LL{{\cal L}}
\def\tir{{\tilde r}}
\def\Ga{{\Gamma}}
\def\ga{{\gamma}}

\def\bp{{\bar{\partial}}}
\def\bz{{\bar{z}}}

\def\sp#1{{\Sigma_{{#1}+}}}
\def\sm#1{{\Sigma_{{#1}-}}}
\def\Sp{{\Sigma_{+}}}
\def\Sm{{\Sigma_{-}}}
\def\s{{\sigma}}
\def\S{{\Sigma}}
\def\fz{{F_0}}
\def\vF{{\vec F}}
\def\vFc{{\vec F}^c}
\def\vn{{\vec n}}
\def\np{{n_+}}
\def\nm{{n_-}}
\def\sf#1{{\Sigma^{F_0}_{#1}}}
\def\zc{{\check z}}
\def\tw{\tilde}
\def\vtF{{\vec{ \tilde F}}}
\def\tS{{\widetilde{\Sigma}}}
\def\hS{{\hat \Sigma}}
\def\hF{{\hat F}}
\def\tA{{\tilde A}}
\def\hA{{\hat A}}
\def\tN{{\tilde N}}
\def\hN{{\hat N}}
\def\hz{{\hat z}}
\def\ti{{\tilde i}}
\def\tj{{\tilde j}}
\def\tir{{\tilde r}}
\def\tI{{\tilde I}}
\def\parl{{(}}
\def\parr{{)}}
\def\tis{{\tilde s}}
\def\bts{\bar{\tilde \sigma}}
\def\hDe{{\hat \Delta}}
\def\nsp{{\sigma_{0+}}}
\def\nsm{{\sigma_{0-}}}

\def\Id{{I'}}
\def\Jd{{J'}}
\def\Kd{{K'}}
\def\vFc{{\vec F^c}}

\def\hbb{{\bar \hbar}}
\def\hbot{\frac{\hbar}{2}}
\def\hbbot{\frac{\bar \hbar}{2}}

\newcommand{\be}{\begin{equation}}
\newcommand{\ee}{\end{equation}}
\newcommand{\bea}{\begin{equation} \begin{aligned}}
\newcommand{\eea}{\end{aligned} \end{equation}}
\newcommand{\nn}{\nonumber}

\newcommand{\bln}{\begin{align}}
\newcommand{\eln}{\end{align}}
\newcommand{\bst}{\begin{split}}
\newcommand{\est}{\end{split}}
\newcommand{\ben}{\begin{enumerate}}
\newcommand{\een}{\end{enumerate}}

%
\catchline{}{}{}{}{}
%

\title{Recent developments in\\
2d $\mathcal{N} = (2,2)$ supersymmetric gauge theories}

\author{Daniel S. Park}

\address{NHETC and Department of Physics and Astronomy\\
Rutgers University, Piscataway, NJ 08855-0849, U.S.A.\\
dspark {\rm at} physics.rutgers.edu}

\maketitle


\begin{abstract}
We review recent developments in two-dimensional
$\mathcal{N}=(2,2)$ supersymmetric gauge theories focusing on the implementation and applications of localization techniques.

\end{abstract}

\ccode{PACS numbers:}

\tableofcontents

\section{Introduction}

Two-dimensional gauge theories with $\cN=(2,2)$ supersymmetry became a topic of intensive research following the poineering work Ref. \citen{WittenGLSM}. While gauge fields do not have propagating degrees of freedom in two dimensions, there is still some rich physics at play. In particular, a 2d gauge theory becomes strongly coupled in the infra-red (IR), opening the possibility for it to flow to a non-trivial fixed point. Ref. \citen{WittenGLSM} not only explains how certain 2d $\cN=(2,2)$ gauge theories---gauged linear sigma models (GLSMs)---flow to non-linear sigma models (NLSMs) of K\"ahler manifolds at intermediate IR scales, but also explains how to understand their IR fixed points.%
\footnote{While Ref. \citen{WittenGLSM} focused on abelian gauge theories, the results were generalized to non-abelian gauge theories in subsequent work \cite{WittenGr,DonagiSharpe,HoriTong}.}
The former fact implies that we are able to compute physical observables of non-linear sigma models, which in turn have geometric interpretations, via gauge theory computations. Furthermore, the fact that 2d $\cN=(2,2)$ gauge theories can be used to engineer $\cN = (2,2)$ superconformal field theories (SCFTs) in the IR also entails that gauge theories can be used to compute observables of type IIA or IIB string compactifications.

The utility of supersymmetric field theories is that there are often exactly computable quantities. One such quantity is the partition function of a supersymmetric theory on a manifold given that the theory has enough supersymmetry. This is because with a certain amount of supersymmetry, the full path integral yielding the partition function can reduce to a integral over a finite dimensional space. This is called {\it localization}, whose first application to quantum gauge theory appeared in the seminal work Ref. \citen{Witten:1988ze}, which studies 4d $\cN = 2$ $SU(2)$ super-Yang-Mills theory. Note that when the manifold is curved, the Lagrangian of a theory on that manifold is not completely determined by its flat space Lagrangian---there is an ambiguity regarding terms that vanish in the flat-space limit. If one chooses the right curvature terms, there exists a supersymmetry, the supercharge operator of which we denote $\cQ$, that is preserved.

In the case of Ref. \citen{Witten:1988ze}, the supercharge in question satisfies $\cQ^2 = 0$. Let us write the partition function $Z_M$ of the theory on a manifold $M$ schematically as
\be
Z_M = \int [d \Phi] \exp\left( -S[\Phi]\right) \,,
\ee
where $\Phi$ denotes all the configuration of fields in the theory. In this review, we exclusively work in the case where $M$ is Euclidean. The expectation value of a $\cQ$-exact operator vanishes, i.e.,
\be
\int [d \Phi] \exp\left( -S[\Phi]\right)
\{\cQ,\cO]= 0 
\ee
given that the measure of integration and action on the manifold is supersymmetric, i.e.,
\be
[\cQ, S ] = 0\,.
\ee
Thus $Z_M$ can be computed also by the path integral
\be\label{loc int}
Z_M = \int [d \Phi] \exp(-S[\Phi]-t\acomm{\cQ}{V})
\ee
for any $t$. The idea of localization is to choose a $\cQ$-exact action $S_\text{loc}=\acomm{\cQ}{V}$ that is bounded below by zero. Upon taking $t$ to infinity, the integral \eq{loc int} localizes onto the locus in field space where $S_\text{loc} = 0$, which we call the {\it localization locus}. The localization locus may have several different components, which we label by $c$. The final integral is then given by
\be\label{loc result}
Z_M = \sum_c \int_{X_c} [dx] \exp(-S[\Phi_x]) Z_\text{1-loop} [\Phi_x]
\ee
where $X_c$ is the component of the localization locus labeled by $c$, and $x$ are coordinates of the localization locus, whose number one would aim to make finite. $S[\Phi_x]$ is the classical action evaluated at the field configuration $\Phi_x$ parametrized by $x$. $Z_\text{1-loop}$ is the one-loop determinant of all the fields around that particular field configuration.

The supercharge studied in Ref. \citen{Witten:1988ze} is topological, in that the stress-energy tensor of the theory is $\cQ$-exact, and thus the partition function of the theory on the manifold is independent of the metric. There are two distinct topological supersymmetric backgrounds for 2d theories with $\cN=(2,2)$ supersymmetry \cite{WittenGW}, coined the $A$-model and the $B$-model \cite{WittenTop}. While these backgrounds have been mostly employed to compute physical observables in non-linear sigma models, localization computations using the $A$-model supercharge have been carried out to compute gauge theory correlators in Ref. \citen{MorrisonPlesser}.

The localization scenario described above can be slightly generalized to incorporate $\cQ$ such that $\cQ^2 = J$, where $J$ is some symmetry of field space. Oftentimes, $J$ is associated to an isometry of the space-time manifold $M$. Given that $[\cQ,S]=0$, we can repeat the argument above by using $V$ that satisfies
\be
[J,V] = 0
\ee
to localize the path integral in question \cite{Witten:1992xu,Moore:1997dj,Moore:1998et}. It would be remiss to go on without emphasizing that the $\Omega$-background \cite{Nekrasov:2002qd,Nekrasov:2003rj} used to compute the exact effective K\"ahler potential of $\cN = 2$ supersymmetric gauge theories fall in this category.

We note that the modern notion of {\it (curved) supersymmetric backgrounds}, studied and developed in Refs. \citen{SF}, \citen{DFS} and \citen{DF}, can be understood as expectation values of fields (including the auxiliary fields) of a supergravity multiplet on manifold $M$.%
\footnote{Also see Refs. \citen{Adams:2011vw, Klare:2012gn}.}
While the expectation values of the fields do not necessarily satisfy equations of motion since gravity is not dynamical in this setting, they are required to preserve some amount of supersymmetry. After the supersymmetry equations are solved, the gravity multiplet may be coupled to a supersymmetric field theory to yield a supersymmetric Lagrangian which in turn may be used to compute the partition function of the theory. We note that in this approach, a field theory must have off-shell supersymmetry for it to couple to the supersymmetric background. A systematic classification of supersymmetric backgrounds in two dimensions coming from the $\cN=(2,2)$ new-minimal supergravity multiplet has been carried out in Ref. \citen{CC}.%
\footnote{See Refs. \citen{GHKSST,BCRR} for subsequent developments.}

The current approach to computing exact partition functions is thus to first pick a supergravity multiplet, classify supersymmetric backgrounds with the given multiplet, use localization to compute the partition function and interpret the result. While pointing out every major development in this direction of research is outside the scope of this review, we have collected a small list of work, starting from the work of Pestun computing the exact partition function of $\cN=2$ gauge theories on the four-sphere \cite{Pestun1}, that have appeared leading up to computations \cite{BC, DGLL,GG,BEHT1,BEHT2, SugishitaTerashima, HondaOkuda, HoriRomo, KLY, BZ, CCP, Benini:2016hjo} of the same spirit being carried out in two dimensions, as well as some early higher-dimensional computations that are directly related to 2d localization \cite{Kapustin:2009kz, Jafferis:2010un, Hama:2010av, Hosomichi:2012ek, Kim:2009wb, Pestun:2009nn, Imamura:2011su, Hama:2011ea, Krattenthaler:2011da, Gomis:2011pf, Kapustin:2011jm, Benini:2011nc, Ito:2011ea, Alday:2012au, Closset:2013sxa,Closset:2016arn}.

Localization computations for 2d $\cN=(2,2)$ gauge theories took off after Refs. \citen{BC,DGLL} computed the exact partition function of the gauge theories on the ``round sphere." Exact partition functions (and/or correlators) on the flat torus \cite{GG,BEHT1,BEHT2}, the hemisphere \cite{SugishitaTerashima, HondaOkuda, HoriRomo}, $\mathbb{RP}^2$ \cite{KLY}, the $\Om$-deformed sphere \cite{BZ,CCP} and the $A$-twisted higher-genus Riemann surface \cite{Benini:2016hjo}%
\footnote{The partition function of a gauge theory on the $A$-twisted higher-genus Riemann surface can be obtained by dimensionally reducing a three-dimensional computation, which also appeared in Ref. \citen{Closset:2016arn}.}
were computed subsequently. While the result of the localization computation for certain backgrounds, such as the $A$-twist or the flat torus, already had well-established expectations, results of some---such as the round sphere partition function for theories with conformal fixed points \cite{JKLMR,GL,GHKSST}---turned out to have surprising physical interpretations. Meanwhile, the meaning of some quantities---such as correlation functions on the $\Om$-deformed sphere---still remain to be understood.

What makes these backgrounds useful is that the path integral can be localized onto relatively unsophisticated saddles, whose moduli space is finite-dimensional. Save for the flat torus, the path integral can be localized onto a Coulomb-branch path integral, where the saddles of the path integral are parametrized by zero modes of the sigma-fields---the scalar components in the vector multiplet---in the Cartan subalgebra of the gauge algebra. At these saddles, the chiral fields are forced to vanish, and can be completely integrated out. This should be contrasted to the localization computation of Ref. \citen{MorrisonPlesser}, where the saddles used are vortex configurations where both the chiral fields and the gauge fields take on non-trivial expectation values.

The goal of this review is to present the results of supersymmetric localization of 2d $\cN=(2,2)$ supersymmetric gauge theories on supersymmetric backgrounds and to discuss their physical and geometric interpretation. We place emphasis on explaining how to actually carry out the computations, and various subtleties that can arise when attempting to evaluate the partition function.

We choose to focus on three backgrounds: the ``round sphere" or equivalently, the two-sphere with no $R$-flux, the torus, and the equivariant $A$-twist on the sphere ($\Omega$-deformed sphere). The gauge theories we consider are those whose gauge multiplets lie within the basic vector multiplet, and whose matter consist of chiral multiplets. Thus the scope of the review is restricted in two directions---the type of gauge theories we consider,%
\footnote{Exact partition functions of field theories with multiplets other than the standard vector and chiral multiplets have been computed in the literature. Partition functions for theories with twisted chiral matter and/or twisted vector multiplets have been computed in Refs. \citen{GL,Doroud:2013pka,Harvey:2014nha}. Refs. \citen{Nian:2014fma,Benini:2015isa} study theories with semi-chiral multiplets.}
and the type of backgrounds we focus on.

This review is structured as follows. We begin by going over basics of $\cN= (2,2)$ gauge theories we study throughout the review in section \ref{s:basics}. We present the field content and the data that specify the gauge theories and explain important aspects of global symmetries. We also introduce and review three (classes of) model theories---the $\mathbb{CP}^{N_f-1}$ model, the quintic GLSM and $U(N)$ gauge theories with fundamental and anti-fundamental matter---whose partition functions we compute on various supersymmetric backgrounds throughout the course of the review. In section \ref{s:backgrounds}, we review the basic framework for understanding the supersymmetric backgrounds. Supersymmetric backgrounds can be understood as vacuum expectation values of bosonic components of a supergravity multiplet coupled to the gauge theories. We review the $\cN=(2,2)$ new minimal supergravity multiplet and write explicit expressions for the Lagrangian of a gauge theory coupled to a supersymmetric background obtained by giving expectation values to components of this multiplet.

In the following three sections, sections \ref{s:round S2}, \ref{s:torus} and \ref{s:omega S2}, we present localization formulae for the round sphere, the torus, and the sphere with the equivariant $A$-twist, respectively. For each of these backgrounds, we describe the localization locus as well as background expectation values for the vector multiplets coupled to the flavor current multiplets that can be turned on. We present the localization formula for gauge theories, and use it to compute the partition function or correlators of model gauge theories introduced in section \ref{s:basics}. We also explain the interpretation of the partition functions and correlators, when they are known.

In section \ref{s:applications} we summarize some applications of the localization computations. On the side of physical applications, we focus on how localization has been used to verify various dualities
\cite{HoriVafa,HoriTong,Hori:2011pd,BC} proposed for 2d $\cN=(2,2)$ gauge theories. We also review how supersymmetric partition functions and correlators are related to geometric invariants of complex K\"ahler manifolds.

We must acknowledge that we have, inevitably, focused on certain topics more than others. It should be emphasized that this is not for the lack of importance of the topics less covered, but simply due to the inability of the author to due justice to all subjects. We give a brief summary of supersymmetric backgrounds that we have not been able to discuss in detail in section \ref{s:other}. We list some more applications of supersymmetric localization in subsection \ref{ss:more applications}.

\section{2d $\cN=(2,2)$ gauge theories} \label{s:basics}

We begin by describing the gauge theories that we wish to study. While there are a plethora of multiplets that can be utilized to construct field theory Lagrangians \cite{}, we stick to gauge theories whose gauge fields lie in vector multiplets, and whose matter fields form chiral multiplets.

While we relegate expressions for the components of the various multiplets and how the various parameters show up in the Lagrangian of the theory to the next section, let us summarize the data that specify the theories that we study. The ingredients that go into specifying the theory are as follows:
\begin{itemlist}
\item The gauge group of the theory $\mathbf{G}$ with gauge algebra $\fg$. The gauge fields sit in the $\fg$ valued vector multiplet, which we denote $\cV$.
 \begin{itemlist}
 \item We denote the rank of $\bfG$, $\rk(\bfG)$.
 \item The Cartan subalgebra $\fh$ of $\fg$ has $\rk({\bfG})$ generators, which we index by the labels $a,b,\cdots$.
 \item The components of $\cV$, in Wess-Zumino gauge, are given by $(a_\mu,\s,\ts,\lam_\pm,\tl_\pm,D)$.
 \item Twisted chiral and twisted anti-chiral multiplets may be constructed by taking supercovariant derivatives of the vector fields in flat space, whose bottom components are given by $\s$ and $\ts$, respectively:
\be
\Sigma = i D_- \widetilde{D}_+ \cV \,, \qquad
\widetilde{\Sigma} = iD_+ \widetilde{D}_- \cV \,.
\ee
We note that in space-times with Euclidean signature, these two superfields are independent, as opposed to the case when the space-time signature is Lorentzian. 
 \item Throughout the review, we refer to the entries of the Lie-algebra valued bosonic component $\s$ of the gauge multiplet as {\it sigma fields}.
 \end{itemlist}
\item Charged chiral and anti-chiral multiplets $\Phi_i$, $\tPhi_i$ in the representations $\fR_i$, $\overline{\fR}_i$ of $\bfG$.
 \begin{itemlist}
  \item We denote the $U(1)_R$-charge of the chiral multiplet $r_i$. The charges $r_i$ must be quantized for certain supersymmetric backgrounds.
  \item Given the flavor symmetry group $\bfG_f$, we can turn on twisted masses $m_F$ for each factor of the Cartan subalgebra $\bigoplus_F \fu(1)_{f,F}$ of the flavor symmetry group. We denote the flavor charge of the chiral fields under $\fu(1)_{f,F}$, $q_F^i$.
  \item On manifolds with a non-trivial fundamental group, holonomies for the $R$-symmetry or flavor symmetries may be turned on along non-contractible loops.
  \item The components of $\Phi_i$ and $\tPhi_i$ are given by $(\phi_i,\psi_{i,\pm},F_i)$ and $(\tphi_i,\tpsi_{i,\pm},\tF_i)$, respectively.%
\footnote{The placement of the flavor indices $i$ with respect to the field symbols vary throughout the review as to make the equations more readable. We hope that this does not cause the reader any confusion.}
 \end{itemlist}
\item The superpotential $W(\Phi_i)$, which is a function of the chiral fields $\Phi_i$. $W$ must be invariant under gauge and flavor symmetry, and must have charge-two under the $U(1)_R$ symmetry.
\item The gauge algebra $\fc$ of the center $\bfC$ of the gauge group is either empty, or is a direct sum of abelian components, which we index by $I=1,\cdots,n$:
\be
\fc = \bigoplus_{I=1}^n \fu(1)_I \,.
\ee
Here, $n \leq \rk({\bfG})$, the equality being saturated when the gauge group is abelian. We introduce the linear twisted superpotential
\be
\hW = {1 \ov 2}\sum_I t_I \s_I \,,
\ee
where $\s_I$ is the complex scalar component of the $I$th abelian vector multiplet. We may write this superpotential in a basis-invariant form as
\be
\hW (\s) = {1 \ov 2} t(\s)
= {1 \ov 2} \sum_I t_I \, \text{tr}_I \s \,,
\ee
where $t$ is a complex vector in the dual of the Cartan subalgebra $\fc_\bC^* \subset \fh_\bC^*$, and the parenthesis denotes the canonical pairing between elements of $\fh^*$ and $\fh$. The parameters $t_I$ are complexified Fayet-Iliopoulos (FI) parameters, whose real part encodes the theta-angle\cite{Coleman} of $\fu(1)_I$. We often write:
\be
t_I = i\xi_I + {\theta_I \ov 2\pi} \,.
\ee
\end{itemlist}
Before we move on, let us be a bit pedantic and review the difference between an $R$-symmetry and a flavor symmetry. $R$-symmetry is a global symmetry under which supercharges carry charge---it follows that components lying in the same multiplet have different representations under the $R$-symmetry. Flavor symmetry, in the context of supersymmetric theories, are global symmetires that commute with supersymmetry---it is not part of the supersymmetry algebra. Thus the flavor charges of all fields in a given chiral multiplet are equivalent.

Classically, in the absence of twisted masses, the $R$-symmetry group is given by $U(1)_R \times U(1)_A$, given that the superpotential $W$ of the theory has $U(1)_R$-charge 2, as we have assumed. We denote the first and second factor the vector and axial $R$-symmetry group, respectively. The $U(1)_R$-charges of the components $(\phi_i,\psi_{i,-},\psi_{i,+},F_i)$ and $(\tphi_i,\tpsi_{i,-},\tpsi_{i,+},\tF_i)$ of the chiral and anti-chiral multiplets are given by $(r_i,r_i-1,r_i-1,r_i-2)$ and $(-r_i,-r_i+1,-r_i+1,-r_i+2)$, while the $U(1)_A$ charges are given by $(0,1,-1,0)$ and $(0,-1,1,0)$. The flavor charges, on the other hand, are given by $(q^i_F,q^i_F,q^i_F)$ and $(-q^i_F,-q^i_F,-q^i_F)$, respectively. Meanwhile, the vector and axial $R$-symmetry charges of the vector multiplet components $(a_\mu,\s,\ts,\lam_-,\lam_+,\tl_-,\tl_+,D)$ are given by $(0,0,0,1,1,-1,-1,0)$ and $(0,2,-2,1,-1,-1,1,0)$, respectively. Note that the twisted masses can be understood as giving a vacuum expectation value to the scalar components of background vector fields coupling to the flavor symmetries. Thus the twisted masses break $U(1)_A$ symmetry explicitly.

While a classical $U(1)_R$ symmetry group is still preserved in the quantum theory, a classical $U(1)_A$ symmetry can be broken by anomalies. Thus when we refer to $R$-charge or $R$-symmetry without further explanation, it should be understood that we are referring to the vector $R$-symmetry. Whether the axial $R$-symmetry of a theory exists in the quantum theory is crucial for determining the properties for its IR fixed point. If the symmetry group $U(1)_A$ is unbroken in the quantum theory, the gauge theory is expected to flow to a superconformal fixed point, as both left and right moving $R$-symmetries, which are part of the superconformal algebra, stay intact. As noted above, the axial $R$-symmetry may be broken by twisted masses classically, or in the absence of twisted masses, broken by mixed anomalies. The mixed anomalies are computed by fermion loop diagrams with two vertices---the axial $R$-symmetry current on one vertex, and the flavor or gauge current on the other. While the mixed anomaly between the $R$-symmetry current and the flavor current can be avoided when the background vector field coupling to the flavor current is not turned on, the mixed anomaly with the gauge current is unavoidable. Nevertheless, a discrete subgroup of $U(1)_A$ can be shown to survive quantum mechanically. We describe these subgroups in relevant examples, which we now present.

Throughout the review, we use three model gauge theories to illustrate how to apply the various localization formulae. The three theories are the following:
\begin{romanlist}
\item The $\bC\bP^{N_f-1}$ model \cite{WittenGLSM,D'Adda:1978kp}.
	\begin{romanlist}
	\item Gauge group: $U(1)$.
	\item Charged matter: $N_f$ chirals $\Phi_i$ with gauge charge 1 and vanishing $U(1)_R$-charge.
	\item $\hW = \ha t \sigma$. $W = 0$. 
     \item $U(1)_A$ symmetry broken to $\bZ_{2N_f}$. 
	\end{romanlist}
\item The quintic GLSM \cite{WittenGLSM}.
	\begin{romanlist}
	\item Gauge group: $U(1)$.
	\item Charged matter: 5 chirals $\Phi_i$ with gauge charge 1 and $U(1)_R$-charge 0, and 1 chiral $P$ with gauge charge $-5$ and $U(1)_R$-charge 2.
	\item $\hW = \ha t \sigma$. $W = PG(\Phi_i)$, $G$ is a homogenous polynomial of degree 5. 
     \item $U(1)_A$ symmetry is unbroken.
	\end{romanlist}
\item $U(N)$ theory with $N_f$ fundamental and $N_a$ antifundamental matter.
	\begin{romanlist}
	\item Gauge group: $U(N)$
	\item Charged matter: $N_f$ fundamental chiral fields $Q_F$, labeled by $F \in [N_f] := \{ 1, \cdots, N_f\}$, and $N_a$ antifundamental chiral fields $\widecheck{Q}_A$, labeled by $A \in [N_a]$. We set the $U(1)_R$-charges $r_F$ and $\rc_A$ to be arbitrary for now.%
	\footnote{Only for this theory are the chiral fields labeled by two different indices $F$ and $A$. As noted before, the index $i$ is used to label all chiral fields, and the index $F$ is used to label elements of the Cartan subalgebra of the flavor symmetry group for all other cases.}
     \item Without loss of generality, we assume $N_f \geq N_a$. We also assume $N_f \geq N$, so that there exists a supersymmetric ground state.
	\item $\hW = \ha t \,\, \text{tr} \,\sigma$. $W$ is a generic gauge invariant polynomial of $U(1)_R$-charge 2, which must be a function of the mesons $\widecheck{Q}_A Q_F$.
     \item When all twisted masses are turned off, $U(1)_A$ symmetry unbroken for $N_f = N_a$; broken to $\bZ_{2(N_f-N_a)}$ for $N_f \neq N_a$.
	\end{romanlist}
\end{romanlist}
We turn on generic twisted masses for the matter in the $U(N)$ theory unless stated otherwise. The flavor symmetry of the theory, in the absence of twisted masses and assuming all $R$-charges are equal, is given by $S[U(N_f) \times U(N_a)]$, whose Cartan subgroup is given by $U(1)^{N_f + N_a -1}$. We choose the conventions where the fundamental fields of $U(N)$ transform as an anti-fundamental of the $U(N_f)$ component, and the antif-fundamental fields of $U(N)$ transform as fundamental fields of $U(N_a)$. The rank of the Cartan subgroup being $(N_f + N_a -1)$, we may turn on as many twisted masses. We choose the more convenient route of turning on $N_f + N_a$ twisted masses, which we denote $s_F$ and $\cs_A$, and identify them under the equivalence relation $(s_F,\cs_A) \equiv (s_F +s , \cs_A +s)$ \cite{BC,BPZ,GLF}. We also can turn on flavor holonomies in a similar manner when we put the gauge theory on $T^2$.

Before going on further, let us comment on the RG flow of these theories. We first discuss theory (i) in some detail, and briefly comment on theories (ii) and (iii). Much of what we discuss can be found in Refs. \citen{WittenGLSM}, \citen{WittenGr}, \citen{DonagiSharpe}, \citen{NS1} and \citen{NS2}. We follow the exposition of Ref. \citen{BPZ} when discussing theory (iii).

Theory (i) flows to a $\mathbb{CP}^{N_f-1}$ sigma-model in the intermediate IR regime. Let us explain what we mean by the intermediate IR regime. The gauge coupling $\mathrm{e}_0$ of a two-dimensional gauge theory has the dimension of mass. Thus the gauge theory becomes strongly coupled at energy scales $\Lambda$ with
\be
\Lambda < {\mathrm{e}_0} \,.
\ee
Meanwhile, there is a second coupling in the game when the gauge group has abelian components, namely the FI parameters. The FI parameters are classically marginal, but can flow---the beta function for the FI parameters are one-loop exact \cite{WittenGLSM} and is given by
\be\label{beta}
\beta = -{b_0 \ov 2 \pi i} \,, \qquad
b_0 = \sum_i \sum_{\rho \in \Lambda_{\fR_i}} \rho \,,
\ee
where we take the view that $\beta$, as $t$, is a vector in $\fh^*_\bC$---in fact, $i \beta$ must be a vector in $i \fc^* \subset \fh_\bC^*$. It proves to be useful to define the vector $b_0$ as defined for future purposes. For theory (i), it follows that $\beta =-N_f/(2 \pi i)$. Thus the effective FI coupling at scale $\Lambda$ is given by
\be
q(\Lambda) =
\left( {\Lambda_\text{UV} \ov \Lambda} \right)^{N_f} q_\text{UV} \,,
\ee
where we have defined
\be
q = e^{2 \pi i t} \,,
\ee
for the FI parameter $t$. We thus find that $q$ becomes large when
\be
\Lambda < {\Lambda_\text{UV} |q_\text{UV}|^{1/N_f} } \,.
\ee
Now we assume that $q_\text{UV}$ is small enough---i.e., the imaginary part of $t$ is positive and very large---so that
\be
|q_\text{UV}| \ll \left( \mathrm{e}_0 \ov \Lambda_\text{UV} \right)^{N_f} \,.
\ee
We define the intermediate IR scale to be when $\Lambda$ is between the two scales where the gauge coupling becomes strongly coupled, and when $q$ becomes small:
\be\label{intermediate}
\Lambda_\text{UV} |q_\text{UV}|^{1/N_f}  \ll \Lambda \ll \mathrm{e}_0 \,.
\ee
In this regime, the effective theory is given by a two-dimensional sigma model into a $\mathbb{CP}^{N_f-1}$ manifold, whose complexified K\"ahler parameter is given by
\be
t(\Lambda) = {1 \ov 2 \pi i}\ln q(\Lambda) \,.
\ee
The K\"ahler parameter is a single complex number, as the second homology group of $\mathbb{CP}^{N_f-1}$ has rank-one. $q$ being small in the regime implies that the imaginary part $\xi$ of the K\"ahler parameter is large and positive, or ``very positive". Only when the imaginary part of $t$ is very positive is the sigma model controllable, since the non-perturbative effects are suppressed by the large volume of the target space.

In the far IR, when
\be\label{far IR}
\Lambda \ll {\Lambda_\text{UV} |q_\text{UV}|^{1/N_f} }  \,,
\ee
the sigma-model is no longer reliable---the theory becomes massive. To be more concrete, there are $N_f$ ground state vacua, and all the fluctuations around a given ground state are massive with masses much larger than $\Lambda$. This can be argued by showing that the extrema of the quantum twisted superpotential $\hW(\sigma)$ are given by the $N_f$ roots of
\be
\s^{N_f} = q
\ee
and that the ground states are reliably represented by these roots. By carefully reinstating all the scales in place, it can be shown that all fluctuations around these vacua have masses of order
\be
|q_\text{UV}|^{1/N_f}\Lambda_\text{UV} \gg \Lambda \,.
\ee

Theory (ii) is quite interesting in two aspects. First, the FI parameter $t$ is exactly marginal and the theory flows to an interacting superconformal theory in the IR. The variable $t$ parametrizes the twisted chiral conformal manifold. Second, depending on the value of $t$, the effective IR theory that describes the theory at length scales where the effective gauge coupling becomes large, takes on different guises. When $|q| \ll 1$, i.e., when the imaginary part of $t$ is very positive, the effective theory is described by a sigma model into the famous quintic Calabi-Yau (CY) threefold. This is often called the {\it large-volume limit} of the SCFT. We should note that the target space of an NLSM that is superconformal has a vanishing canonical class, i.e., is Calabi-Yau. Meanwhile, when the imaginary part of $t$ is very negative, the effective theory is a Landau-Ginzburg orbifold theory. It is quite surprising that the extended conformal manifolds of these two seemingly very different theories are the same, and that in fact the two theories have an interpretation as different ``phases" of the same superconformal theory \cite{WittenGLSM}. Note that when discussing theory (ii), we do not mention an intermediate IR scale, as the only energy scale in play is the scale $\mathrm{e}_0$, where the effective gauge coupling becomes large. As before, in the ``geometric phase" where the theory flows to an NLSM into the quintic threefold, $t$ is identified with the complexified K\"ahler parameter of the manifold. Meanwhile, the coefficients of the superpotential, which are also exactly marginal deformations of the SCFT, parametrize the complex structure moduli space of the manifold.

Theory (iii) has many moving parts. Let us first discuss the theory where all twisted masses are turned off, and the $R$-charges are set to zero. The beta function for the FI parameter is given by $\beta = (N_a-N_f)/(2 \pi i)$, and the theory flows to a superconformal theory in the IR only when $N_f = N_a$. When $N_a = 0$, and the imaginary part of $t$ is taken to be very positive, the theory flows to an NLSM of the Grassmannian $\Gr(N,N_f)$, which is the space of complex $N$-planes in $\bC^{N_f}$, in the intermediate IR regime. Each anti-fundamental matter can be interpreted as a copy of the tautological bundle $S$ of the Grassmannian. Thus in general, when the imaginary part of $t$ is taken to be sufficiently large, the theory flows to an NLSM of the bundle $S^{\oplus N_a} \rightarrow \Gr(N,N_f)$ in the IR. For $N_f = N_a$, the theory flows to the NLSM of the manifold $S^{\oplus N_f} \rightarrow \Gr(N,N_f)$ for both limits $|q| \ll 1$ and $|q| \gg 1$. The two manifolds, however, are not equivalent, and are related by a flop. In particular, the complexified K\"ahler parameter of the manifold for $|q| \ll 1$ is identified with $t$, while the K\"ahler parameter of the manifold obtained for $|q| \gg 1$ is identified with $-t$.

The elements of the cohomology of the manifold $X := S^{\oplus N_a} \rightarrow \Gr(N,N_f)$ are represented by gauge-invariant polynomials of the sigma fields, as explained in Ref. \citen{WittenGr}.%
\footnote{While all the elements of the cohomology of $S^{\oplus N_a} \rightarrow \Gr(N,N_f)$ are represented this way for theory (iii) with all $R$-charges set to zero, this is not the case in general. For example, in the geometric phase of theory (ii), there are 204 elements of the third cohomology of the quintic threefold that cannot be represented using the sigma fields.}
The operators built out of the sigma fields are elements of the quantum cohomology ring \cite{Vafa:1991uz} of the manifold, and their correlation functions encode the Gromov-Witten invariants \cite{Gromov,DSWW,WittenGW} of these manifolds. Now the flavor symmetry of the gauge theory translates into isometries of the NLSM one obtains in the IR. Given the existence of the $U(1)^{N_a+N_f-1}$ isometry of manifold $X$, one could ask about the equivariant version \cite{AtiyahBott,Witten2D} of the quantum cohomology \cite{Vafa:1991uz} and Gromov-Witten invariants \cite{GiventalEquivariant} of $X$. These can be computed by turning on the twisted masses $s_F$ and $\cs_A$ for the isometries. Below the energy scale of the twisted masses, the theory quickly flows to a theory of $\left( \begin{smallmatrix} N_f \\ N\end{smallmatrix} \right)$ isolated massive vacua \cite{NS1,NS2}. The correlators of this theory, however, encode the equivariant quantum cohomology of the original manifold $\cM$ with equivariant parameters $s_F$ and $\cs_A$.

\section{Supersymmetric backgrounds}
\label{s:backgrounds}

In this section, we discuss general aspects of supersymmetric backgrounds that we can use to localize gauge theory on. We also write down the Lagrangians of  gauge theories in the supersymmetric backgrounds. We follow the approach of Ref. \citen{CC}, where the supersymmetric backgrounds are obtained by turning on expectation values for fields in the $\cN = (2,2)$ new minimal supergravity multiplet. We note that all the formulae for the supersymmetric backgrounds and Lagrangians in this section are borrowed from Ref. \citen{CC}.%
\footnote{While we focus on gauge theories with chiral matter in this work, it is straightforward to couple any $\cN=(2,2)$ field theory whose flat space Lagrangian can be written in superspace, such as a non-linear sigma model, to these supersymmetric backgrounds \cite{CC}. Curvature couplings of 2d $\cN=(2,2)$ NLSMs have also been studied in Ref. \citen{Jia:2013foa}.}
 		
As explained in the introduction, the supersymmetric backgrounds can be understood as vacuum expectation values (VEVs) of components of the $\cN = (2,2)$ new minimal supergravity multiplet. We utilize the bosonic components of this multiplet denoted by
\be\label{bosonic fields}
g_{\mu \nu} \,, \quad
A_\mu \,, \quad
C_\mu \,, \quad
\tC_\mu
\ee
in Ref. \citen{CC}. Obviously, $g_{\mu \nu}$ is the background metric, while $A_\mu$ is the vector field that couples to the $R$-current. $C_\mu$ and $\tC_\mu$ are vector fields that couple to the current associated to the complex central charge of the supersymmetry algebra. As in Ref. \citen{CC}, it is useful to introduce the field strengths
\be
\cH = -i \epsilon^{\mu \nu} \p_\mu C_\nu \,, \quad
\tH = -i \epsilon^{\mu \nu} \p_\mu \tC_\nu \,.
\ee
Following Ref. \citen{CC}, we use ``$R$" to denote the {\it opposite} of the scalar curvature in this section. Given that these components are turned on, the Lagrangian of the theory is given by
\be \mathscr{L}_\text{gauge}+\mathscr{L}_\text{chiral}
+\mathscr{L}_{W} + \mathscr{L}_{\hW} \,.
\ee
Each term is written explicitly as follows.

The term $\mathscr{L}_\text{gauge}$ is the gauge-kinetic term:%
\footnote{When the gauge algebra can be decomposed into a direct sum of sub-algebras, we may assign to each component a separate gauge coupling. The gauge coupling, however, does not play a prominent role in the backgrounds we study, as we see later on.}
\bea
\mathscr{L}_\text{gauge} ={1 \ov \mathrm{e}_0^2}
&\Bigg[
D_1 \ts D_\bo \s +D_\bo \ts D_1 \s + {1 \ov 8} [\s,\ts]^2 \\
&+2i \tl_+ D_1 \lam_+ - 2 i \tl_- D_\bo \lam_-
+ i\tl_- [\s,\lam_+] - i \tl_+[\ts,\lam_-] \\
&+{1 \ov 2} \left( 2if_{1 \bo} +{1 \ov 2} \tH \s - {1\ov 2} \cH \ts \right)^2
-{1 \ov 2} \left(D +{1 \ov 2} \tH \s + {1\ov 2} \cH \ts \right)^2  \Bigg] \,.
\eea
We note that we use the frame indices $1$ and $\bo$, where the metric is written
\be
ds^2 = 2 g_{z \bar{z}} dz d\bar{z} = e^{1} e^\bo \,.
\ee
Note that by definition, $f_{1 \bo}$ is imaginary. The covariant derivative $D_\mu$ defined to be
\be
D_\mu = \nabla_\mu -ir A_\mu + {1 \ov 2} s \tC_\mu -{1 \ov 2} \widetilde{s} C_\mu
\ee
where $\nabla_\mu$ is the covariant derivative including the metric and gauge connections. Here, $r$ is the $R$-charge of the field, while $s$ and $\widetilde{s}$ are the complex central charges of the field. We note that $s = \widetilde{s} =0$ for fields in the vector multiplet, but these are non-vanishing for fields in the chiral multiplet when twisted masses are turned on. We discuss this point further later on.

The kinetic term for the chiral fields can be written in the form
\be
\mathscr{L}_\text{chiral}=
{1 \ov \mathrm{g}^2} \sum_i \mathscr{L}_i
\ee
where
\bea
\mathscr{L}_i &=2D_1 \tphi^i D_\bo \phi^i + 2D_\bo \tphi^i D_1 \phi^i - \tF^i F^i
+ 2 i \tpsi^i_+ D_1 \psi^i_+ - 2i \tpsi^i_- D_\bo \psi^i_- + \tphi^i D \phi^i \\
&-\left( {r_i \ov 4}R -{1 \ov 2} \cH \tes_i - {1 \ov 2} \tH s_i \right) \tphi^i \phi^i
+ \tphi^i \left(\tes_i -\ts -{r_i \ov 2} \tH \right) \left( s_i - \s - {r_i \ov 2}\tH \right) \phi^i \\
&+{1 \ov 2} \tphi^i [\s,\ts] \phi^i
 + i\tpsi^i_+ \left( \tes_i-\ts -{r_i \ov 2} \tH \right) \psi^i_-
 -i\tpsi^i_- \left( s_i -\s -{r_i \ov 2} \cH \right) \psi^i_+ \\
 &+ i \sqrt{2} (\tpsi^i_+ \tl_- -\tpsi^i_- \tl_+) \phi^i
 + i \sqrt{2} \tphi^i(\lam_+ \psi^i_- -\lam_- \psi^i_+) \,.
\eea
Here, all the fields in the vector multiplet can be understood as being matrices in the $\fR_i$ representation of $\fg$, and that the indices of the fields are contracted accordingly. When we wish to turn on real twisted masses only, we may take
\be
s_i = \tes_i = \sum_F s_F q^i_F \,,
\ee
where $s_F$ are twisted masses for the flavor symmetry $\mathfrak{u}(1)_{f,F}$, and $q^i_F$ is the flavor charge of $\Phi_i$.

It turns out to be more useful to allow more elaborate flavor backgrounds by turning on supersymmetric expectation values for background vector multiplets coupled to the flavor symmetry. This is particularly useful if one wishes to ultimately promote a subgroup of the flavor symmetry group into a gauge symmetry. To do so, we need to set all the central charges $s$ and $\tes$ to zero, and couple the chiral fields to the background flavor vector multiplet $\cV_F$. Now turning on a constant real twisted mass $s_F$ for a flavor symmetry is equivalent to turning on the supersymmetric expectation values
\be
s_F = \tes_F  \,, \quad
a_{F,\mu} = - {i \ov 2} s_F \tC_\mu + {i \ov 2} \tes_F C_\mu\,, \quad
D_F= -{1 \ov 2} s_F \tH - {1 \ov 2} \tes_F \cH
\ee
for the fields in the vector multiplet $\cV_F$. There are more general supersymmetric configurations of $\cV_F$ that may be turned on depending on the supersymmetric background the theory is coupled to. We explore such configurations further in the subsequent sections.

The superpotential terms of the Lagrangian are determined by functions $W$ and $\tW$ of $\Phi_i$:
\bea
\mathscr{L}_W = F^i \p_i W(\phi_i) + \psi^i_- \psi^j_+ \p_i \p_j W(\phi_i) +
\tF^i \tp_i \tW(\tphi_i) + \tpsi^i_- \tpsi^j_+ \tp_i \tp_j \tW(\tphi_i) \,. 
\eea
for the superpotential $W(\Phi_i)$ of the theory. While $W$ and $\tW$ do not have to related in Euclidean signature, we restrict ourselves in the case where $\tW$ is related to $W$ by
\be
\tW(\overline{\phi_i}) = \overline{W(\phi_i)} \,,
\ee
where the bar denotes complex conjugation.

We mostly consider twisted superpotentials that are linear in the field strength multiplet of the abelian factors of the gauge group. The twisted superpotential terms are then given by:
\be
\mathscr{L}_{\hW} = - \sum_I \xi_I \text{tr}_I D
+ i \sum_I {\theta \ov 2\pi} \text{tr}_I (2i f_{1 \bo}) \,, 
\ee
where $f_{1 \bo}$ is the field strength of the gauge field. For future reference, let us note that for generic twisted superpotential functions $\hW$ and $\thW$ of twisted superfields $\Om_i$ with components $(\om^i, \eta^i_\pm, G^i)$, the twisted superpotential terms are given by
\bea
\mathscr{L}_{\hW} = &G^i \p_i \hW(\om_i)
+ \eta^i_- \teta^j_+ \p_i \p_j \hW (\om_i)
- i \tH \hW (\om_i) \\
&+\tG^i \tp_i \thW(\tom_i)
- \teta^i_- \eta^j_+ \tp_i \tp_j \thW (\tom_i)
+ i \cH \thW (\tom_i)
\,.
\eea
As before, we restrict ourselves to cases when the function $\thW$ is defined with respect to $\hW$ such that
\be
\thW(\overline{\om_i}) = \overline{\hW(\om_i)} \,.
\ee
We note that the components of the twisted chiral field-strength multiplet constructed from the vector multiplet has components
\be
(\s,\sqrt{2}\lam_-,-\sqrt{2}\tl_+, iD -2f_{1\bo}+i\tH\s) \,.
\ee

The supersymmetric backgrounds are found by asking which expectation values of \eq{bosonic fields} preserve some supersymmetry. The supersymmetry transformations of fields in the supergravity multiplet can be found in Ref. \citen{CC}, as well as a systematic classification of supersymmetric backgrounds. Now in all the supersymmetric backgrounds we discuss in this review, the Lagrangians $\mathscr{L}_\text{gauge}$, $\mathscr{L}_\text{chiral}$ and $\mathscr{L}_{W}$ are exact, in that they can be written as $\{\cQ,V \}$ for a preserved supercharge $\cQ$.%
\footnote{While most of the supersymmetric backgrounds for $\cN=(2,2)$ studied can be understood as vacuum expectation values of the new minimal supergravity multiplet, the torus partition function we study in section \ref{s:torus} is an exception, and should be understood as a background of a $\cN=(0,2)$ supergravity multiplet. There, a background gauge field coupling to the left-moving $R$-current, which does not exist in the $\cN=(2,2)$ supergravity multiplet studied here, is turned on.}
In fact, we use
\be
\left\{\cQ,{1 \ov \mathrm{e}_0^2}V_\text{gauge} +
{1 \ov \mathrm{g}^2}V_\text{chiral}\right \} = 
\mathscr{L}_\text{gauge} +\mathscr{L}_\text{chiral}
\ee
as the localizing Lagrangian of equation \eq{loc int} to localize the path integral to the Coulomb branch, i.e., we take $\mathrm{e}_0, \mathrm{g} \ra 0$. Thus one might naively expect that the partition function is only dependent on parameters of the twisted superpotential $\tW$. This is almost true---the one-loop determinants of the theory depend on the quadratic fluctuations around the localization locus, and these can depend on additional parameters of the theory, such as the $R$-charge, twisted masses or background fluxes of flavor symmetries, as we see in specific examples later on. This is consistent with the fact that such parameters appear in the supersymmetry algebra of the localizing supercharge being used.

\section{The round sphere}
\label{s:round S2}

In this section, we review the round sphere partition function (or equivalently, the partition function on the $S^2$ with no $R$-flux) first computed in Refs. \citen{BC} and \citen{DGLL}.  The background supergravity fields for the round sphere partition function are given by \cite{CC}:
\be
ds^2 = {4 \ov (1+|z|^2)^2} dz d \bar z \,, \quad
A_\mu =0 \,, \quad
\cH = {i} \,, \quad
\tH = {i} \,,
\ee
where we have set the radius of the sphere to $1$. We note that the gauge field $A_\mu$ that couples to the $R$-symmetry current is set to vanish in this background.

The partition function can be made to localize onto the {\it Coulomb branch locus} where all the components of the chiral fields vanish and the fields $\s$, $\ts$, $D$ and $f_{1 \bo}$ take on the constant values
\be\label{saddles}
\s = \hsig - {i \ov 2} \fm \,, \quad
\ts = \hsig +{i \ov 2} \fm \,, \quad
D = -i \hsig \,, \quad
2if_{1 \bo} = -{\fm \ov 2} \,. 
\ee
Here $\hsig$ and $\fm$ are taken to be elements of the Cartan subalgebra $i\fh$. We often write the components of $\hsig$ and $\fm$ explicitly, i.e.,
\be
\hsig =  \hsig_a T^a \,, \qquad
\fm = \fm_a T^a \,,
\ee
where $T^a$ are generators of the Cartan subalgebra. As we have set the radius of the sphere to be the unit of length, $\fm$ is identified with the magnetic flux through the sphere, and must be GNO quantized \cite{GNO}. Thus the path integral of the theory on the round sphere localizes onto a sum over the magnetic flux $\fm$, and a finite dimensional integral over the continuous real parameters $\hsig_a$.

As discussed previously, we may turn on complex twisted masses in these backgrounds by turning on components of the background flavor vector multiplet $\cV_F$. These complex masses are turned on by giving a vacuum expectation value to the scalar components $s_F$ and $\tes_F$ of $\cV_F$. By supersymmetry, the vector field $a_{F,\mu}$ and auxiliary field $D_F$ of the multiplet must also take the following expectation values:
\be
s_F = \re(s_F) - {i \ov 2} \fm_F \,, \quad
\tes_F = \re(s_F) +{i \ov 2} \fm_F \,, \quad
D = - i \re(s_F) \,, \quad
2if_{F,1 \bo} =- {\fm_F \ov 2} \,. 
\ee
We have denoted the imaginary part of $s_F$, $\fm_F$ to emphasize that $\fm_F$ is a background magnetic flux for the flavor symmetry. This flux must be quantized so that $q^i_F \fm_F \in \bZ$ for all chiral fields $\Phi_i$. As before, we denote the complex twisted masses for the multiplets $\Phi_i$ and $\tPhi_i$,
\be
s_i = q^i_F s_F \,, \quad
\tes_i = q^i_F \tes_F \,.
\ee

The round sphere partition function is given by%
\footnote{Operators and defects may be inserted in the path integral, although we do not explore this possibility in this section. The insertion of vortex defects in this background has been studied in Ref. \citen{Hosomichi:2015pia}.}
\be\label{ZS2}
Z_{S^2} =
{1 \ov |\cW|}\sum_\fm \int \left( \prod_a {d\hsig_a \ov 2 \pi} \right)  \, Z_\fm^{cl}(\hsig) \,
Z_{\cV,\fm}^{1\ell}(\hsig) \, \prod_i Z_{i,\fm}^{1\ell} (\hsig) \,,
\ee
where the sum $\fm$ runs over all GNO quantized fluxes. $|\cW|$ denotes the order of the Weyl group of the gauge algebra. $Z^{cl}_\fm(\hsig)$ is the classical action evaluated at the saddles:%
\footnote{We note that it is not entirely accurate to call this piece the classical piece, in that the renormalized real FI parameter $\xi_I$ at the scale of the radius of the sphere should be plugged into equation \eq{classical} \cite{BC, DGLL}.}
\bea\label{classical}
Z^{cl}_\fm(\hsig) &= \exp \left( - 4 \pi i \sum_I \xi_I \tr_I (\hsig)
+ i \sum_I \theta_I \tr_I(\fm) \right) \\
&= e^{-2 \pi \sum_I (t_I \tr_I \s - \bar{t}_I \tr_I \ts)}
= e^{-2 \pi t(\s) + 2 \pi  \bar{t}(\ts)}\,.
\eea
As we see as we go on, the expressions become more elegant, once the variables $\s$ and $\ts$ in equation \eq{saddles} are used. The other factors of the integrand come from the one-loop determinant of the vector and chiral multiplets. $Z_{\cV,\fm}^{1\ell}(\hsig)$ comes from the vector multiplets:
\be\label{Zvect}
Z_{\cV,\fm}^{1\ell}(\hsig) = \prod_{\alpha > 0}
\left(- \alpha(\hsig)^2 - {\alpha(\fm)^2 \ov 4}\right)
= \prod_{\alpha > 0}
\Big(-\alpha(\s)\alpha(\ts) \Big)
\ee
where the product runs over the positive roots $\alpha >0$ of the Lie algebra $\fg$.%
\footnote{There is an additional minus sign on each of the factors of the product in equation \eq{Zvect} compared to Refs. \citen{BC,DGLL}, which has been correctly accounted for in Ref. \citen{HoriRomo}.}
The one-loop factor coming from integrating out $\Phi_i$ and $\tPhi_i$ is given by
\be
Z_{i,\fm}^{1\ell} (\hsig) = \prod_{\rho \in \Lambda_{\fR_i}}
{\Ga\left({r_i \ov 2} - i\rho(\hsig)-{\rho(\fm) \ov 2} - i s_i \right)
\ov
\Ga\left(1-{r_i \ov 2} + i\rho(\hsig)-{\rho(\fm) \ov 2} + i \tes_i \right)}
= \prod_{\rho \in \Lambda_{\fR_i}}
{\Ga\left({r_i \ov 2} - i\rho(\sig) - i s_i \right)
\ov
\Ga\left(1-{r_i \ov 2} + i\rho(\ts)+ i \tes_i \right)} \,,
\ee
where $\Lambda_{\fR_i} \subset i \fh^*$ denote the set of weights of representation $\fR_i$.

The integral \eq{ZS2} is a real integral, in that the contour of integration for $\hsig_a$ is along the real line, when all the twisted masses $s_i = \tes_i$ are real. In many interesting cases, the integrand of \eq{ZS2} may have poles along the real axis. The correct way to deal with those cases is to first shift the $R$-charge of the chiral fields whose one-loop determinant is responsible for the poles by a small positive amount $\delta r$.%
\footnote{Depending on the properties one wants to preserve, this may not be possible. To be more concrete, one might want certain superpotential terms to be present, and thus may want to impose linear constraints on the supercharges of the various chiral fields in the theory. In this case, one needs to shift the $R$-charges in a manner consistent with the linear constraints, such that all gauge invariant polynomials of the chiral fields have positive $R$-charge.}
After this deformation, the integral \eq{ZS2} will not have any poles along the real axis, and may be evaluated. The desired partition function may then be obtained by taking the limit $\delta r \ra 0$.%
\footnote{While the partition function $Z_{S^2}$ often has a well defined $\delta r \ra 0$ limit, it may as well diverge. For example, such divergences are present when the theory flows to an NLSM of a non-compact Calabi-Yau manifold in the IR. Even in such cases, it has been shown that information about the IR theory is encoded in the leading singular terms with respect to $\delta r$ \cite{PS,Honma:2013hma}.}

To be concrete, let us write the explicit matrix integrals for the round sphere partition function for theories (i), (ii) and (iii) introduced in section \ref{s:basics}. For theory (i), the partition function is given by
\be
\sum_{\fm \in \bZ} \int d \hsig e^{- 4 \pi i \xi \hsig + i \theta \fm}
\left( {\Ga\left( - i \hsig -{\fm \ov 2}\right)
\ov
\Ga\left(1+ i \hsig -{\fm \ov 2} \right)}\right)^{N_f} \,.
\ee
For theory (ii), it is given by
\be
\sum_{\fm \in \bZ} \int d \hsig e^{- 4 \pi i \xi \hsig + i \theta \fm}
\left( {\Ga\left( - i \hsig -{\fm \ov 2}\right)
\ov
\Ga\left(1+ i \hsig -{\fm \ov 2} \right)}\right)^{5}
{\Ga\left(1+ 5i \hsig +{5 \ov 2} \fm\right)
\ov
\Ga\left(-5i \hsig +{5 \ov 2} \fm \right)} \,.
\ee

For theory (iii), we write the partition function in a particular way that turns out to be quite useful for multiple purposes. To do so, we introduce some notation, following Ref. \citen{BPZ}. Note that the Cartan subalgebra of $\mathfrak{u}(N)$ is given the diagonal entries and thus the saddles are given by the diagonal matrices
\be
\hsig = \text{diag}(\hsig_a) \,, \qquad
\fm = \text{diag}(\fm_a) \,.
\ee
with $\fm_a \in \bZ$.
We define the ``packaged" variables
\bea
\s_{a,+} &= i\s_a = i \hsig_a + {\fm_a \ov 2}\,, &
\s_{a,-} &= i\ts_a = i \hsig_a - {\fm_a \ov 2} \\
\S_{F,\pm} &= i\re(s_F) \pm {\fm_F \ov 2} + {r_F \ov 2} \,, &
\cS_{A,\pm} &= i\re(\cs_A) \pm {\widecheck{\fm}_A \ov 2}-{\rc_A \ov 2}\,,
\eea
and the traces
\be
\S_{\pm} = \sum_a \s_{a,\pm} \,.
\ee
We introduce the following differences to condense our notation. We distinguish the various entries of the differences by their indices as follows:
\bea
\S^a_{b\pm} &= \s_{a\pm}-\s_{b\pm}\,, &
\S^a_{F\pm} &= \s_{a\pm}-\S_{F\pm}\,, &
\S^a_{A\pm} &= \s_{a\pm}-\cS_{A\pm}\,, \\
\S^{F_1}_{F_2\pm} &= \S_{F_1\pm}-\S_{F_2\pm}\,, &
\S^F_{A\pm} &= \S_{F\pm}-\cS_{A\pm}\,, &
\S^{A_1}_{A_2\pm} &= \cS_{A_1\pm}-\cS_{A_2\pm}\,.
\eea
The partition function for theory (iii) may now be succinctly written as
\bea\label{UN}
{e^{i\varphi_N} \ov N!}
\sum_{\fm \in \bZ^N}
\int \left( \prod_a  {d \s_a \ov 2\pi} \right)
q_+^{\S_+} q_-^{\S_{-}}
\prod_{a<b} \S^a_{b+} \S^a_{b-}
\prod_F {\Ga(-\S^a_{F+}) \ov \Ga(1+\S^a_{F-})}
\prod_A {\Ga(\S^a_{A+}) \ov \Ga(1-\S^a_{A-})}
\eea
with the phase $\varphi_N = N(N-1) \pi/2$. We have previously introduced the exponential of the complexified FI parameter $q = e^{2 \pi i t}$. $q_\pm$ are related to $q$ by
\be
q_+ = e^{-i \pi (N+1)} q \,, \quad
q_- = e^{i \pi (N+1)} \bar{q} \,.
\ee
The contour of integration of the integral \eq{UN} needs to be commented on, as we have turned on flavor fluxes. By carefully reviewing the prescription for the contour, one should be able to convince oneself that the contour should be taken such that all the poles coming from the one-loop determinant of the fundamental fields should be positioned below the $\hsig_a$-contour, while those coming from the determinant of the anti-fundamental fields should be positioned above the contour. In other words, the contour should be taken such that it divides the two classes of poles.

When the beta function of an FI parameter $\xi$ is negative or zero, the asymptotics of the integrand becomes such that we may deform the contour of integration to the lower-half plane for each $\hsig_a$ coupled to $\xi$ linearly, once we take $\xi$ to be very positive. The final integral then can be written as a sum of residues of poles of the integrand over poles lying below the contour of integration. When the beta function is positive or zero, we can similarly deform the contour of integration to the upper-half $\hsig_a$ plane once we take $\xi$ to be very negative.

For the $U(N)$ theories at hand, we can deform the contour of integration to the lower-half plane and pick up the poles at
\be\label{poles}
\s_{a+}= \S_{F_a+}+n_+ \,, \qquad
\s_{a-}= \S_{F_a-}+n_- \,,
\ee
for $n_\pm \in \bZ_{\geq 0}$. The partition function then decomposes into the form
\be\label{decomposition}
\sum_{\vF \in C(N,N_f)} Z_0^\vF Z^\vF_+ Z^\vF_- \,.
\ee
Here, $C(N,N_f)$ is the set of $N$-tuples of integers $F_a$ such that
\be
1 \leq F_1 < F_2 < \cdots < F_N \leq N_f \,.
\ee
The component labeled by $\vF$ comes from picking up precisely the poles \eq{poles} for components of $\vF$. 

The decomposition \eq{decomposition} has a beautiful interpretation, as the result of the localization computation on a different set of saddles, which is often called the {\it Higgs branch locus}. A vector $\vF \in C(N,N_f)$ labels a Higgs sector of the theory, where the fundamental fields $F_1, \cdots, F_N$ take vacuum expectation values. In order for the fields to do so, the sigma fields must take vacuum expectation values such that
\be\label{Higgs bulk}
\s \equiv \text{diag}(s_{F_a}) \,, \qquad
\ts \equiv \text{diag}(\tes_{F_a}) \,,
\ee
where the symbol ``$\equiv$" in this equation is used to indicate equivalence up to conjugation. The Higgs branch saddles are such that in the bulk of the sphere the fundamental fields $F_1, \cdots, F_N$ take constant VEVs and the sigma fields take constant values \eq{Higgs bulk}. One then needs to sum over the point-like vortices and anti-vortices localized at the two poles of the sphere, in which the fundamental fields $F_1, \cdots, F_N$ are turned on. The factor $Z_0^\vF$ is the one-loop determinant of the fields around the constant field configuration, while $Z_+^\vF$ and $Z_-^\vF$ denote the {\it vortex} and {\it anti-vortex partition functions} \cite{Shadchin}, coming from summing over all the vortex configurations localized at the two poles of the sphere. Quite amazingly, the sphere partition function allows one to compute the vortex partition function reliably without addressing the vortex moduli space.

While we do not write the explicit expression for $Z^\vF_0$, which can be found in many places in the literature \cite{BC,DGLL,BPZ,GLF}, let us note that the vortex partition functions are given by
\be
Z^\vF_+ = Z_\text{v}^{\vF} (\S_{F_+};\cS_{A+};q) \,, \qquad
Z^\vF_- = Z_\text{v}^{\vF} (\S_{F_-};\cS_{A-};(-1)^{N_a-N_f}\bar{q}) \,,
\ee
where we have used the shorthand notation $Z_\text{v}^{\vF} = Z_\text{v}^{\vF,\cO=1}$ for the function
\bea\label{VPF}
&Z_\text{v}^{\vF,\cO} (\S_{F};\cS_{A};q) = \\
&\sum_{n \geq 0} q^n
 \sum_{|(n_a)|=n} \cO(\Sigma_{F_a}+n_a) 
\cdot \prod_{a=1}^N {\prod_{A=1}^{N_a}(\S^{F_a}_A)_{n_a} \ov
\prod_{b=1}^N(-\S^{F_a}_{F_b} -n_a)_{n_b}
\prod_{b'=1}^{N'} (-\S^{F_a}_{F^c_{b'}}-n_a)_{n_a}}
\eea
defined for any symmetric polynomial $\cO$ of $|\vF| =N$ variables. We have denoted $(n_a)$ to be $N$-tuples of non-negative integers, and $|(n_a)| := \sum_a n_a$. $\vFc$ is a $N' = (N_f-N)$-tuple in $C(N_f,N_f-N)$, whose elements are given by the complement of $\vF$ with respect to $[N_f]$. $(a)_n$ is the Pochammer symbol, given by $\Ga(a+n)/\Ga(a)$.

When the theory flows to a $\cN = (2,2)$ superconformal fixed point in the IR, the sphere partition function $Z_{S^2}$ has a beautiful interpretation first conjectured in Ref. \citen{JKLMR}, and proven in Refs. \citen{GL,GHKSST}. The FI parameters $t_I$ of the gauge theory are marginal deformations of the superconformal theory, and thus span the conformal manifold of twisted chiral couplings of the IR theory. This conformal manifold has a K\"ahler metric \cite{Seiberg:1988pf,Periwal:1989mx,Cecotti:1991me}, which can be identified with the Zamolodchikov metric \cite{Zamolodchikov:1986gt}. The sphere partition function can be related to the K\"ahler potential of this metric by
\be
Z_{S^2} (t_I, \bar{t}_I) = e^{-K(t_I, \bar{t}_I)}\,.
\ee
Note that the K\"ahler potential $K$ is determined up to shifts
\be
K(t_I, \bar{t}_I) \ra K(t_I, \bar{t}_I) + f(t_I ) + \overline{f(t_I)} \,,
\ee
where $f$ is a holomorphic function of the parameters $t_I$. These come from local counterterms in the field theory \cite{CC,GHKSST}, which vanish in flat space. Thus, it is appropriate to think of the partition function $Z_{S^2}$ as a section of a bundle over the twisted chiral conformal manifold, rather than a function.

When the superconformal theory in the IR is an NLSM with a target manifold $X$, the FI parameters $t_I$ can be identified with the K\"ahler parameters of $X$, and the conformal manifold with its extended K\"ahler moduli space. Thus, in this case, $Z_{S^2}$ turns out to encode sophisticated geometric invariants, more about which we discuss in section \ref{s:applications}.

We end the discussion of the round sphere partition function by noting that while $Z_{S^2}$ has an elegant interpretation when the gauge theory flows to a superconformal fixed point in the IR, its meaning is not entirely clear when this is not the case. It would be interesting to understand the significance of $Z_{S^2}$ when the IR theory is not superconformal.

\section{The torus}
\label{s:torus}

We now study the torus partition functions, computed in Refs. \citen{GG,BEHT1} and \citen{BEHT2}. We first review their results, which assumes that the gauge group is connected. We then slightly extend these results to the case when the gauge group contains a discrete factor in section \ref{ss:discrete}.

The supersymmetric background for the torus is quite simple---the metric is flat, and a flat connection for the gauge field $A^L_\mu$, which couples to the left-moving $R$-current of the theory, may be turned on. We can think of the torus as being obtained by quotienting the complex plane with complex coordinate $w$ with respect to two independent shifts such that
\be\label{shifts}
w \sim w + 1 \sim w + \tau \,,
\ee
where $\tau$ is the complex structure of the torus. The flat connection for the left-moving $R$-symmetry, which we denote $U(1)_L$, is parametrized by the complex holonomy
\be
z = \oint_t A^L - \tau \oint_s A^L \,,
\ee
where $t$ and $s$ denote the temporal, and spatial cycles on the torus. In order to turn on a holonomy $z$ for the $U(1)_L$-charge, all twisted masses for the chiral multiplets must be turned off, as their introduction completely breaks the left-moving $R$-symmetry classically. Meanwhile, flat connections of background flavor symmetries may be turned on, parametrized by
\be
\cu_F = \oint_t a_F - \tau \oint_s a_F \,.
\ee
The parameters $z$ and $u_F$ both lie on a torus with complex structure $\tau$, as they have the periodicities \eq{shifts}. Following Refs. \citen{GG,BEHT1,BEHT2}, we define the exponentiated parameters:%
\footnote{In this section and this section only do we use the variable $q$ to denote the exponentiated complex structure of the torus. This is not to be confused with the exponentiated FI parameters, which the torus partition function does not depend on.}
\be
q = e^{2 \pi i \tau} \,, \quad
y = e^{2 \pi i z} \,, \quad
\cx_F = e^{2\pi i \cu_F} \,.
\ee
We note that when the left-moving $R$-symmetry of the theory is broken into a discrete subgroup $\Gamma$ of $U(1)_L$ by anomalies, the partition function is not well defined unless $y$ is restricted to be an element of $\Gamma$.

The path integral for the torus partition function localizes onto the space of flat connections, parametrized by the holonomies of the connection around the two cycles of the torus. As with the background flavor and $R$-symmetry connections, this can be packaged into a complex parameter for each element of the Cartan subalgebra of the gauge group, which we denote $u_a$, following Refs. \citen{GG,BEHT1,BEHT2}. After an integration by parts, the path integral can be shown to be
\be\label{torus}
Z_{T^2} = {1 \ov |\cW|}
\int_{\cC} \prod_a{ d u_a \ov 2 \pi i}  \, Z^{1\ell}_\cV( u) \, \prod_i Z^{1\ell}_i (u) \,,
\ee
with a middle-dimensional integration cycle $\cC$ in the $\rk(\bfG)$-complex dimensional torus, which we denote $\tfT$. The torus $\tfT$ can be identified with the smooth cover of the space of flat connections on $T^2$.  As before, $|\cW|$ is the order of the Weyl group of the gauge group. We note that there is no sum over distinct topological sectors when the gauge group is connected. The one-loop determinants coming from the vector multiplet, and the chiral multiplet $i$ are given by
\be
Z^{1\ell}_\cV \!=\! \left( {2 \pi \eta(q) \ov \theta_1 (q,y^{-1}) }\right)^{\rk(\bfG)}
\!\!\!
\prod_\alpha {\theta_1 (q, x^\alpha) \ov \theta_1 (q,y^{-1}x^\alpha)} \,, \quad
Z^{1\ell}_i \!=\!
\prod_{\rho \in \Lambda_{\fR_i}} {\theta_1 (q, y^{r_i/2-1}x^\rho \cx^{q^i}) \ov \theta_1 (q, y^{r_i/2}x^\rho \cx^{q^i})} \,,
\ee
where $r_i$ is the $U(1)_R$-charge of the chiral multiplet. As before, $\alpha$ runs over the roots of the gauge algebra, while $\Lambda_{\fR_i}$ denotes the weights of the representation $\fR_i$ of $\bfG$. We have introduced the notation such that
\be
x^\rho = e^{2\pi i \rho(u)} \,, \quad
\cx^{q^i} = e^{2\pi i q^i_F \cu_F} \,.
\ee
The eta function and theta function have the expansions
\be\label{eta theta}
\eta(q) = q^{1\ov 24} \prod_{n=1}^\infty (1-q^n) \,, \quad
\theta_1 (q,y) = -i q^{1 \ov 8} y^{1 \ov 2} \prod_{k=1}^\infty
(1-q^k)(1-yq^k)(1-y^{-1}q^{k-1})
\ee
when the arguments are small. Note that these functions are multivalued with respect to the variables $q$ and $y$, and should really be viewed as functions of $\tau$ and $z$ for $y = e^{2\pi i z}$. Following Refs. \citen{BEHT1,BEHT2}, we write the function as $\theta_1 (\tau|z)$ when we want to emphasize this fact. $\theta_1 (\tau|u)$ does not have poles with respect to $u$, while it has zeros at $u = n + m \tau$. We find that
\be
{\theta_1 (\tau|u-z) \ov \theta_1(\tau|u)} \sim
{y^m \theta_1(q,y^{-1})
\ov 2\pi \eta(q)^3  (u-n-m\tau)}
\ee
around this point. It is also worth noting that
\be
\theta_1(\tau|z+n+m\tau) =
(-1)^{n+m} e^{-2\pi i m z - i \pi m^2 \tau } \theta_1(\tau|z) \,.
\ee

Now the most non-trivial part of the equation \eq{torus} lies in the determination of the contour of integration $\cC$. Note that the integrand of the integral \eq{torus} is holomorphic at a generic point on the complex $\rk(\bfG)$-dimensional torus. Thus, the integral \eq{torus} boils down to a sum of residues of the integrand. The singularity of the factors of the integrand lie along the hyperplanes:
\be
H_\cI = \left\{~ u ~:~ Q_\cI (u) + {r_\cI \ov 2} z + q^\cI_F \cu_F \equiv 0 \quad (\text{mod $\bZ + \tau \bZ$})~ \right\} 
\ee
where $\cI$ labels all the charged components of the chiral multiplets in the theory. To be more concrete, the label $\cI$ belongs to the set
\be\label{cI def}
\cI \in
 \{~ (\rho,i) ~:~\rho \in \Lambda_{\fR_i}~ \} \,.
\ee
Recall that we use the index $i$ to label the chiral multiplets throughout this review. Then we find that
\be\label{charges etc}
(Q_\cI,r_\cI,q^\cI_F) =
(\rho,r_i,q^i_F) \,.
\ee

Now let us define the set of codimension-$\rk(\bfG)$ singularities of the integrand of \eq{torus}, $\tfT^*_\sing$. For any $u_* \in \tfT^*_\sing$, there are $s \geq \rk(\bfG)$ hyperplanes $H_{\cI_1} , H_{\cI_2}, \cdots, H_{\cI_s}$ intersecting at $u_*$. We use $\bfQ(u_*)$ to denote the charges associated to those hyperplanes:
\be
\bfQ(u_*) = \{ Q_{\cI_1}, \cdots, Q_{\cI_s }\}\,.
\ee
The torus partition function \eq{torus} can then be written as
\be\label{torus new}
Z_{T^2} = \sum_{u_* \in \tfT^*_\sing}
\JKRes{u = u_*}\left[\bfQ(u_*), \eta \right] \,
\om_{T^2} (u) \,,
\ee
where we have defined the differential form
\be
\om_{T^2}(u) :=
Z^{1\ell}_\cV( u) \, \prod_i Z^{1\ell}_i (u)
\, du_1 \wedge \cdots \wedge du_{\rk(\bfG)} \,.
\ee
In order to evaluate this formula we must review the definition of the Jeffrey-Kirwan residue (JK residue) \cite{JK1995,1999math......3178B,2004InMat.158..453S}.%
\footnote{Refs. \citen{BEHT1,BEHT2} re-introduced Jeffrey-Kirwan residues into the physics literature, which has been showing up in localization computations in diverse dimensions and backgrounds since \cite{HKY,BZ,CCP,Benini:2016hjo}. A generalization of the JK residue, coined the Jeffrey-Kirwan-Grothendieck residue, appears in localization computations of half-twisted correlators of $\cN = (0,2)$ GLSMs \cite{CGJS}.}
A vector $\eta \in i\fh^*$ must be introduced to compute these residues. While the individual residues depend on the choice of $\eta$, the final sum, which yields the partition function, is independent of this vector.

After shifting the position of the pole to the origin, the definition of the JK residue boils down to defining the residue of a rational $\rk(\bfG)$-form $\Om$ that can have poles along the hyperplanes $Q_{\cI}(u) = 0$ for $Q_{\cI} \in \bfQ(u_*)$. The space of such differential forms $R_{\bfQ(u_*) }$ can be understood as a graded vector space over the complex numbers. There is a subspace $S_{\bfQ(u_*)}$ of this vector space, spanned by
$\left( \begin{smallmatrix}|\bfQ(u_*)| \\ \rk(\bfG)\end{smallmatrix}\right)$ $\rk(\bfG)$-forms:
\be
\om_{\calS} = \prod_{Q_{\cI} \in \calS} {1 \ov Q_{\cI} (u)}  du_1 \wedge \cdots \wedge du_{\rk(\bfG)}
\ee
where $\calS$ is an arbitrary subset of $\bfQ(u_*)$ with $\rk(\bfG)$ distinct elements. For $\om_{\calS}$, we define the JK residue to be
\be
\JKRes{u =0} \left[\bfQ(u_*), \eta \right] \, \om_\calS= 
\begin{cases}
{1 \ov |\det (S)|} & \text{when} \quad \eta \in \text{Cone}(\calS) \\
0 & \text{when} \quad \eta \notin \text{Cone}(\calS) \,,
\end{cases}
\ee
where $\text{Cone}(\calS)$ is the cone in $i\fh^*$ spanned by the elements of $\calS$, and $\det(S)$ is the determinant of the $\rk(\bfG) \times \rk(\bfG)$ matrix of charge vectors $Q_\cI \in \calS$. This defines the JK residue for any element of the vector space $S_{\bfQ(u_*)}$. There is a natural projection $\pi$ from $R_{\bfQ(u_*) }$ to $S_{\bfQ(u_*)}$, which is the analogue of extracting the $z^{-1}$ term of the Laurent series of a rational function $f(z)$ at $z=0$. Then, the JK residue of any $\Om \in R_{\bfQ(u_*)}$ is defined to be the JK residue of the projection $\pi(\Om)$ of $\Om$ into $S_{\bfQ(u_*)}$ \cite{CCP,1999math......3178B,2004InMat.158..453S}.

In order for $\om_{T^2} (u)$ to be a single-valued differential form on the torus $\tfT$, it must be invariant under the shifts $u_a \ra u_a+1$ and $u_a \ra u_a + \tau$ for each index $a$. One can check that while $\om_{T^2} (u)$ is automatically invariant under the shifts $u_a \ra u_a+1$, invariance under the latter shift is not guaranteed with the holonomy $z$ for the left-moving $R$-symmetry turned on. The failure of $\om_{T^2}$ to be invariant under this shift reflects the $U(1)_L$-anomaly of the theory.

Let us touch on a rather technical point, before moving on further. When the charges of the hyperplanes meeting at a singularity are contained within a half-space of $i \fh^*$, the hyperplane arrangement is said to be projective. When the arrangement of the intersecting hyperplanes at some $u_*$ is not projective, the JK residue is not well defined. To resolve such a situation, one must judicially shift parameters such as flavor charges and $R$-charges to split-up the non-projective singularity in to projective ones, compute the partition function, and take the limit where the parameters are set to their initial values.

Let us now write down some expressions for the torus partition function of the model theories (i), (ii) and (iii), all of which can be found in the original works \cite{BEHT1,BEHT2}. For theory (i), we have
\be\label{T1 torus}
\om_{T^2}(u) = {2 \pi \eta(q)^3 \ov \theta_1 (q,y^{-1})}
\left( {\theta_1 (q,y^{-1}x) \ov \theta_1 (q,x)} \right)^{N_f}
du \,.
\ee
It is simple to verify that
\be
\om_{T^2}(u + \tau) = y^{N_f}\om_{T^2}(u) \,,
\ee
and $\om_{T^2}$ is not single-valued on $\tfT$ unless $y$ is an element of the multiplicative group $\bZ_{N_f}$. In order to employ the formula \eq{torus new}, we must turn on $y$ such that $y \neq 1$, since when $y=1$, $\om_{T^2}$ diverges, and is not well-defined. Obviously, the torus partition function computed by the formula \eq{torus new} vanishes for $y^{N_f}=1$ such that $y \neq 1$, since we may take $\eta=-1$ in formula \eq{torus new} and find that none of the poles of \eq{T1 torus} contribute to the partition function. Meanwhile, it is argued in Ref. \citen{BEHT1} that $Z_{T^2} = N_f$ for $y=1$.%
\footnote{Alternatively, since we have turned off the Wilson line for the left-moving $R$-symmetry, we may introduce twisted masses and use the localization formula for the $A$-twist on the torus to find that $Z_{T^2} = N_f$ \cite{Benini:2016hjo}.}

This is consistent with the fact that the $\mathbb{CP}^{N_f-1}$ theory has $N_f$ massive vacua represented by the $N_f$ solutions of
\be
\sigma^{N_f} = e^{2 \pi i t} \,,
\ee
where $t$ is the complexified FI parameter of the theory. For this theory, the torus partition function computes a trace of the operator $y^J$, $J$ being the charge of $U(1)_L$, with respect to these ground states. The sigma field $\s$ has unit charge under the action of the left-moving $R$-symmetry. The $k$-th vacuum $\ket{k}$ corresponds to the expectation value
\be
\ket{k}~: \quad
\sigma = e^{2\pi i k /N_f}e^{2 \pi i t/N_f}
\ee
of the sigma field. We see that a generic element of the left-moving $R$-symmetry group maps a vacuum to a non-vacuum. A discrete subgroup of the $R$-symmetry, however, is preserved, since the set of vacua are preserved by a $\bZ_{N_f}$ subgroup. The generator $\om = e^{2\pi i /N_f}$ of this subgroup maps $\ket{k}$ to $\ket{k+1}$. Thus the vacua may be organized in representations of $\bZ_{N_f}$:
\be
\ket{\tilde{k}} = \sum_{k=0}^{N_f-1} e^{-2\pi i \tilde{k} k/N_f}\ket{k} \,.
\ee
It is simple to see that
\be
\om \ket{\tilde{k}} = e^{2\pi i \tilde{k}/N_f}
\ket{\tilde{k}}
\ee
where $\om$ on the left-hand-side of the equation is understood to be an operator, rather than a number. Then the trace over the ground states can be computed explicitly:
\be
\tr \, \om^k = \sum_{\tilde{k}=0}^{N_f-1}
e^{2\pi i k \tilde{k}/N_f}
= \begin{cases}
N_f & \text{when $k \equiv 0~(\text{mod}~N_f)$} \\
0 & \text{otherwise,}
\end{cases}
\ee
which is confirmed by the localization computation.

For theory (ii), we find that
\be\label{T2 torus}
\om_{T^2}(u) = {2 \pi \eta(q)^3 \ov \theta_1 (q,y^{-1})} \cdot
{\theta_1 (q,x^{-5})  \ov
 \theta_1 (q,yx^{-5}) } \cdot
 { \theta_1 (q,y^{-1}x)^5 \ov
\theta_1 (q,x)^5 }  du \,.
\ee
The theory is superconformal, and thus $\om_{T^2}$ is well-defined on $\tfT$ for any value of $y$. We can take $\eta = -1$ in equation \eq{torus new} and find that only residues the second factor of \eq{T2 torus} contribute to the partition function. There are 25 poles at $x = e^{2\pi i (z + k + l \tau)/5}$ for $0 \leq k, l \leq 4$ whose residues can be summed up to give:
\be\label{ZT2 quintic}
Z_{T^2} = {1 \ov 5} \sum_{k,l=0}^4 y^{-l}
\left( {\theta_1 (q,e^{2\pi i (-4z + k + l)/5})
\ov \theta_1 (q,e^{2\pi i (z + k + l)/5})} \right)^5 \,.
\ee

$\om_{T^2} (u)$ for theory (iii) is given by
\bea
\om_{T^2} (u) = &
{1 \ov N!} \left( {2 \pi \eta(q)^3 \ov \theta_1(q,y^{-1})} \right)^N
\left( \prod_{a \neq b}
{\theta_1(q,x_ax_b^{-1}) \ov \theta_1(q,y^{-1}x_ax_b^{-1})}\right) \\
&\prod_a \left(
\prod_F {\theta_1 (q,y^{r_F/2-1}x_a \breve{x}_F^{-1}) \ov
\theta_1 (q,y^{r_F/2}x_a \breve{x}_F^{-1})}
\prod_A {\theta_1 (q,y^{\rc_A/2-1}x_a^{-1} \cx_A) \ov
\theta_1 (q,y^{\rc_A/2}x_a^{-1} \cx_A)}
\right)
du_1 \wedge \cdots \wedge du_N
\eea
where we have turned on holonomies $\breve{x}_F = e^{2\pi i \breve{u}_F}$ and $\cx_A = e^{2 \pi i \cu_A}$ for the flavor symmetry. We refer the reader to the original reference Ref. \citen{BEHT2} for the evaluation of this partition function.

The torus partition function, as computed, has a natural interpretation as the elliptic genus \cite{Pilch:1986en,Witten:1986bf,Alvarez:1987wg}---a weighted sum over the Ramond-Ramond states of the two-dimensional theory
\be
Z_{T^2} = \mathrm{Tr}_{RR} (-1)^F \bar{q}^{H_R} q^{H_L}
y^{J} \prod_F x_F^{K_F} \,,
\ee
where $H_R$ and $H_L$ are the right- and left-moving Hamiltonian operators while $J$ and $K_F$ are the charge operators for the  left-moving $R$-symmetry and the maximal torus of the flavor symmetry, respectively. $F$ is the fermion number. Note that when the left-moving $R$-symmetry is broken to a finite subgroup $\Gamma$ of $U(1)_L$, $y^J$ must be restricted to be an element of $\Gamma$. When $y$ and $x_F$ are all taken to $1$, we arrive at the celebrated Witten index \cite{Witten:1982df}, which counts the (graded) number of ground states of the theory.

When the gauge theory flows to an NLSM in the IR, the elliptic genus has an interpretation as the index of a Dirac operator in the loop space of the target manifold \cite{Witten:1987cg}. This is a geometric invariant that can be computed by integrating a certain elliptic density over the manifold \cite{Kawai:1993jk}. Now for an NLSM that has a (compact) Calabi-Yau threefold as a target space, the elliptic genus is completely determined by the Euler number $\chi_E$ of the threefold \cite{Kawai:1993jk,Keller:2012mr}:
\be
\mathrm{Tr}_{RR} (-1)^F \bar{q}^{H_R} q^{H_L} y^{J}
 = {1 \ov 2} \chi_E \, \phi_{0,3/2} (q,y) =
 {1 \ov 2} \chi_E  (y^{1/2} + y^{-1/2})  + \cO(q) \,.
\ee
Here, $\phi_{0,3/2}$ is a weak Jacobi form, written out explicitly, for example, in Ref. \citen{Keller:2012mr}. We can check that the computation \eq{ZT2 quintic} for the quintic GLSM reproduces the Euler number $\chi_E=-200$ of the quintic threefold by utilizing the expansion \eq{eta theta} and taking the limit $q \ra 0$ :
\be
Z^\text{quintic}_{T^2}|_{q \ra 0} =
-100(y^{-1/2} + y^{1/2}) \,.
\ee

\subsection{Discrete gauge symmetry} \label{ss:discrete}

While we have discussed theories with a connected gauge group up to this point, there are many interesting theories whose gauge group has multiple components. In this section, we discuss the simplest case, when the gauge group factors into a continuous, and a discrete factor:
\be
\bfG = \bfG_\text{cont} \times \mathbf{\Ga} \,.
\ee
In this case, we must sum over all non-trivial principal $\mathbf{\Ga}$ bundles over the manifold $M$ the gauge theory lives on. This can be readily computed by \cite{BundleBook}
\be\label{principal}
\text{Hom}(\pi_1 (M),\bfGa)/\bfGa
\ee
where the quotient is taken with respect to the adjoint action of $\bfGa$. We thus see that the sphere does not have any non-trivial principal $\bfGa$-bundles, while the torus does. The elements of \eq{principal} for the torus can be explicitly written out to be
\be
\{\, (g,h)~ :~ gh= hg \,\}/\sim
\ee
where $(g_1, h_1) \sim (g_2, h_2)$ when $(g_1,h_1) = (g^{-1}g_2 g, g^{-1} h_2 g)$ for some $g \in \bfGa$.

Now the chiral fields $\Phi_i$ in the theory transform as representations of $\bfGa$. Thus for each $g \in \bfGa$, there exists a matrix $\Lam(g)_{ij}$ such that the action of $g$ is given by
\be
g(\Phi)_i = \Lam(g)_{ij} \Phi_j \,.
\ee
Let us introduce the twisted partition function $Z_{T^2} (g,h)$, which is the torus partition function with twisted boundary conditions such that
\be
\Phi_i (w+1)= \Lam(g)_{ij} \Phi_j (w) \,, \quad
\Phi_i (w+\tau)= \Lam(h)_{ij} \Phi_j (w) \,.
\ee
Then the partition function of the theory $Z_{T^2}$ must be given as a weighted sum over the twisted partition functions $Z_{T^2} (g,h)$ for all commuting pairs of $g$ and $h$. To find the correct weights, we can view the gauge theory as a $\bfG_\text{cont}$ gauge theory orbifolded by the global symmetry $\bfGa$. Following Ref. \citen{DHVW1}, we then arrive at
\be\label{orbifold}
Z_{T^2} = {1 \ov |\bfGa|} \sum_{gh = hg} Z_{T^2} (g,h) \,.
\ee

We note that one can do something more interesting, while we do not explore this possibility further here. For $\bfGa$ whose second group cohomology $H^2(\bfGa,U(1))$ is non-trivial, we can consider a theory with discrete torsion which can be identified with a choice of an element $\gamma$ of this cohomology group \cite{VTorsion}. Once the torsion is turned on, the torus partition function is given by
\be
Z_{T^2} = {1 \ov |\bfGa|} \sum_{gh = hg} \epsilon_\gamma (g,h) Z_{T^2} (g,h)
\ee
for some non-trivial phases $\epsilon_\gamma (g,h)$. From the gauge theory point of view, this is equivalent to coupling the $\bfG_\text{cont}$ gauge theory with global symmetry $\bfGa$ to a non-trivial topological field theory \cite{DW}.

Now let us consider two simple examples of gauge theories that flow to a Calabi-Yau threefold in the IR, both $\bZ_5$ orbifolds of the quintic threefold theory, at special points in the complex structure moduli space where the theory has a $\bZ_5$ symmetry. Both theories have
\be
\bfG = U(1) \times \bZ_5
\ee
as their gauge group, with six chiral multiplets as matter: $\Phi_i$ with $U(1)$ charge $1$ and $P$ with $U(1)$ charge $-5$, only differing in their charges under $\bZ_5$. The charges $q_i$ of the chiral multiplets under $\bZ_5$ for the two theories, which we denote (ii)-A and (ii)-B, are listed in table \ref{t:Z5}. Now the superpotential of these theories must be further restricted to be invariant under the given $\bZ_5$ symmetry---this is equivalent to moving to a point in complex-structure moduli space where the threefold has the appropriate $\bZ_5$ isometry.

\begin{table}[h!]
\tbl{Charges of chiral fields of theory (ii)-A and (ii)-B under $\bZ_5$.}
{\begin{tabular}{ccccccc} \toprule
& $\Phi_1$ & $\Phi_2$ & $\Phi_3$ &
 $\Phi_4$ & $\Phi_5$ & $P$ \\ \colrule
(ii)-A & 0 & 1 & 2& 3& 4& 0 \\
(ii)-B & 0&0&0&1&4&0 \\
\botrule
\end{tabular} \label{t:Z5}}
\end{table}

Now in the geometric phase of the IR theory, the action of the $\bZ_5$ group in the IR becomes an orbifolding action of the target manifold. The Euler numbers of the CY manifolds obtained by orbifolding the quintic by the actions of table \ref{t:Z5} can be found in Ref. \citen{Aspinwall:1990xe}, among other places. We note that for (ii)-A, the action is free, i.e., does not have any fixed points, and the resulting theory does not have a twisted sector. This means that all the states of the orbifold theory can be obtained by projecting the states of the quintic theory down to $\bZ_5$-invariant subspace. This results in the Euler number becoming a fifth of the Euler number of the quintic, i.e., $\chi_E = -40$. On the other hand, theory (ii)-B does have fixed points and twisted sector states. Due to these states, the Euler number of the orbifold is given by $\chi_E = -88$.

These Euler numbers can be nicely computed by the torus partition function, using the formula \eq{orbifold}. $\bfGa$ being abelian, the torus partition function can be written as
\bea
Z_{T^2} &= {1 \ov 5} \sum_{r,s=0}^4 Z_{T^2} (e^{2\pi i r/5},e^{2\pi i s/5}) \\ 
&= {1 \ov 5} \sum_{r,s=0}^4 \sum_{u_* \in \tfT^*_\sing}
\JKRes{u = u_*}\left[\bfQ(u_*), \eta \right] \,
\om^{r,s}_{T^2} (u) \,.
\eea
We can readily compute this partition function for theories (ii)-A and (ii)-B.

For (ii)-A, $\om_{T^2}^{r,s}$ is given by
\be
\om_{T^2}^{r,s} = {2 \pi \eta(q)^3 \ov \theta_1 (q,y^{-1})} 
{\theta_1(q,x^{-5}) \ov \theta_1(q,yx^{-5})} 
\prod_{m=0}^{4} {\theta_1(q,y^{-1}x e^{2\pi i m( s- r\tau )/5} )
\ov \theta_1(q,x e^{2\pi i m( s-r \tau )/5} )} du
\ee
We now show that
\be
Z_{T^2}(e^{2\pi i r/5},e^{2 \pi i s/5}) =0
\ee
for $(r,s) \neq (0,0)$. For such $(r,s)$, let us define $\zeta = e^{2\pi i (s-r \tau)/5}$. Now taking $\eta = 1$, we find that there are five potential simple poles located at distinct points $x = \zeta^{-m}$ for $m=0,\cdots,4$. The JK residue at the pole $x = \zeta^{-m}$, however, is given by
\be
{\theta_1 (q,\zeta^{-5m}) \ov 
\theta_1 (q,y\zeta^{-5m})}
\prod_{0 \leq m (\neq k) \leq 4} {\theta_1 (q,y^{-1}\zeta^{m-k}) \ov 
\theta_1 (q,\zeta^{m-k})} = 0 \,,
\ee
since $\zeta^5 = q^{-r}$, $\theta_1 (q,q^{rm}) = 0$ and the factors $\theta_1 (q,\zeta^{m-k})$ are all non-zero when $(r,s) \neq (0,0)$. Hence we find that
\be
Z^\text{(ii)-A}_{T^2} = 
{1 \ov 5} Z^\text{quintic}_{T^2} \,,
\ee
consistent with the claim that the twisted sectors of theory (ii)-A should be empty. Thus it is easy to see that
\be
Z^\text{(ii)-A}_{T^2} |_{q \ra 0} =
{1 \ov 5} Z^\text{(ii)-A}_{T^2} |_{q \ra 0}
= -20 (y^{-1/2} + y^{1/2})
\ee
reproducing the Euler number $\chi_E = -40$ for the orbifold.

Theory (ii)-B, on the other hand, has states in the twisted sector. $\om^{r,s}_{T^2}$ for this theory is given by
\bea\label{iiB}
\om_{T^2}^{r,s} = &{2 \pi \eta(q)^3 \ov \theta_1 (q,y^{-1})} \cdot
{\theta_1(q,x^{-5}) \ov \theta_1(q,yx^{-5})} \cdot
\left( {\theta_1(q,y^{-1}x)
\ov \theta_1(q,x)} \right)^3 \\
& \cdot
 {\theta_1(q,y^{-1}x e^{2\pi i ( s- r\tau )/5} )
\ov \theta_1(q,x e^{2\pi i ( s-r \tau )/5} )}
\cdot
 {\theta_1(q,y^{-1}x e^{8\pi i ( s- r\tau )/5} )
\ov \theta_1(q,x e^{8\pi i ( s-r \tau )/5} )} du \,.
\eea
We may take $\eta =-1$ and evaluate the JK residues at the 25 poles $x = e^{2\pi i (k+l \tau)/5}$ of the second factor of equation \eq{iiB} for each of the 25 twisted partition functions. Taking the $q \ra 0$ limit, we indeed arrive at
\be
Z^\text{(ii)-B}_{T^2} |_{q \ra 0} = -44 (y^{-1/2}+y^{1/2}) \,,
\ee
reproducing the Euler number $\chi_E = -88$.

\section{The $\Omega$-deformed sphere}
\label{s:omega S2}

Let us move on to describing the equivariant $A$-twisted sphere, or the $\Omega$-deformed sphere. This background was studied mainly in Ref. \citen{CCP}, but could be obtained by dimensionally reducing a supersymmetric background on $S^1 \times S^2$ along the $S^1$ direction \cite{BZ}. We follow the exposition of Ref. \citen{CCP}.

The supersymmetric background is given by the expectation values
\be\label{omega S2 bg}
ds^2 = g_{z \bar z}(|z|^2) dz d\bar{z}\,, \quad
A_\mu = {1 \ov 2}\om_\mu\,, \quad
\cH =  {\epsdef \ov 2} \epsilon^{\mu\nu}
\p_\mu V_\nu \,, \quad
\tH = 0\,,
\ee
where $\om_\mu$ is the spin connection of the metric and $V_\mu$ is defined to be the Killing vector
\be
V_\mu =  iz \p_z - i \bar{z} \p_{\bar{z}}
\ee
of the $U(1)$ isometry. The metric can be any smooth metric with the isometery generated by $V_\mu$. The localizing supercharge on this background squares to the generator for the action of the isometry, $\epsdef$ being the equivariant parameter. Thus this background is the two-sphere analogue of the omega deformation \cite{Nekrasov:2002qd,Nekrasov:2003rj} in four dimensions. Note that the background $U(1)$ gauge field coupling to the $R$-charge has unit magnetic flux through the sphere. Thus, in order to couple a theory consistently to the background, all the fields of the theory must have integer $R$-charge.

We take the localizing action to be the standard gauge and chiral kinetic terms.%
\footnote{As in the case of the round sphere, one may choose a localizing action that localizes to a Higgs branch locus \cite{CCP}.}
The saddle of the action is simple in the zero-flux sector---it is given by setting $\s=\overline{\ts}$ to a constant real value with vanishing field strength $2if_{1 \bo}$. Meanwhile, explicit expressions for the supersymmetric field configuration of the saddle points in sectors with non-zero gauge flux have not been obtained. Nevertheless, assuming the existence of such saddles, just enough information to compute the partition function can obtained by utilizing supersymmetry and index theorems. In particular, it can be shown that the bosonic zero modes, which can be identified as the coordinates of the moduli space of saddle points, are given by
\be
\hsig_a = {(\s_a)_S + (\s_a)_N \ov 2} \,,
\ee
where the subscript denotes the value of the field at the south or north pole of the sphere. By supersymmetry, it can be shown that the saddles parametrized by $\hsig_a$ in a given flux sector satisfies
\be\label{sigma saddle}
(\s_a)_N = \hsig_a - {1 \ov 2} \epsdef \fm_a \,,
\quad
(\s_a)_S = \hsig_a + {1 \ov 2} \epsdef \fm_a \,,
\ee
where the $\fm_a$ are the fluxes of the Cartan elements of the gauge group:
\be
{1\ov 2\pi}\int e^1 e^\bo (-2i f_{1 \bo})
= \fm_a T^a \,.
\ee

Recall that we may turn supersymmetric vacuum expectation values of the vector multiplets that couple to the flavor symmetries of the theory. We may thus turn on supersymmetric field configurations of the sigma fields and the gauge fields in the flavor vector multiplets parametrized by $s_F$ and $\fm_F$ so that 
\be\label{tm saddles}
(s_F)_N = s_F - {1 \ov 2} \epsdef \fm_F \,,
\quad
(s_F)_S = s_F + {1 \ov 2} \epsdef \fm_F \,,
\quad
{1 \ov 2\pi}\int (da_F) = {\fm_F} \,,
\ee
where $(s_F)_{S,N}$ are used to denote the vacuum expectation values of the sigma field of the flavor vector multiplet at the south and north poles. We note that unlike in the case of the round sphere, $s_F$ may be any complex number. Setting $\fm_F = 0$ corresponds to turning on an ordinary complex twisted mass.

The integration over these saddles has the extra complication of having gaugino zero modes around them, as was the case for the saddles of the torus partition function studied in the previous section. The partition function can be schematically written as
\be
\sum_{\fm} \int
d\hsig \, d\widetilde{\hsig} \, d\widehat{\lam}
\, d\widetilde{\widehat{\lam}} \,  d\widehat{D}
\, \cZ_\fm (\hsig,\widetilde{\hsig},\widehat{\lam},\widetilde{\widehat{\lam}},\widehat{D})
\ee
where $(\hsig,\widetilde{\hsig},\widehat{\lam},\widetilde{\widehat{\lam}},\widehat{D})$ form a zero-mode multiplet, and the sum over magnetic fluxes $\fm$ is taken. Due to the fact that $\cZ_\fm$ is invariant under the relevant supersymmetries, it can be shown that the integrals, for each $\fm$, reduces to a holomorphic integral for $\hsig_a$ over a middle dimensional contour in $\bC^{\rk(\bfG)}$.

The partition function of a gauge theory coupled to this background generically vanishes. One can nevertheless insert operators at the poles of the sphere to compute expectation values or correlators of operators. While more general operators preserving the supersymmetry can be constructed, we consider the correlation functions of gauge invariant operators constructed using the sigma fields of the vector multiplet. For example, for the $U(N)$ theory, the ring of these operators is generated by
\be
\tr \, \s^k \,, \quad  k=1,\cdots,N \,.
\ee
We note that any such operator $\cO$ can be written as a polynomial of the eigenvalues $\sigma_a$ of $\sigma$ that is invariant under the Weyl group of the gauge group. We often write $\cO(\s_a)$ to denote the polynomial corresponding to the operator $\cO$.

The result of the path integral, with operators $\cO_1$ and $\cO_2$ inserted at the north and south poles, is given by the weighted sum of the JK residues
\bea\label{omega S2}
\vev{ \cO_1 \evon{\rN}\cO_2 \evon{\rS}}_\epsdef
= &{(-1)^{N_*} \ov |\cW|} \epsdef^{-d_\text{grav}} \\
&\cdot \sum_\fm {e^{2\pi i t_s (\fm)} \ov \epsdef^{b_0(\fm)}}
\sum_{\hsig_* \in \tfM^\fm_\sing}
\JKRes{\hsig = \hsig_*} [\bfQ(\hsig_*),\xi^\UV_\eff]
\, \mathbf{I}_\fm (\cO_1, \cO_2) \,,
\eea
of the differential form
\bea
\mathbf{I}_{\fm}  (\cO_1, \cO_2)
=&{1 \ov \epsdef^{\rk(\bfG)}}
 \cdot \cO_1(\hsig_\rN)
\cdot \cO_2(\hsig_\rS)
\, d \hsig_1 \wedge \cdots \wedge d\hsig_{\rk(\bfG)}\\
& \cdot
\prod_{\alpha > 0}
\left( {\a(\hsig_\rN) \a(\hsig_\rS) \ov \epsdef^2} \right)
\prod_i \prod_{\rho \in \Lambda_{\fR_i}}
{\Ga\left({\rho(\hsig_\rN) + s_{i,\rN} \ov \epsdef} + {r_i \ov 2}\right)
\ov
\Ga\left({\rho(\hsig_\rS) + s_{i,\rS} \ov \epsdef} - {r_i \ov 2}+1\right)} 
\,.
\eea
Here we have used the packaged variables
\bea
\hsig_{\rN} &= \hsig - {1 \ov 2} \epsdef \fm \,, &
\hsig_{\rS} &= \hsig + {1 \ov 2} \epsdef \fm \,, \\
s_{F,\rN} &= s_F - {1 \ov 2} \epsdef \fm_F \,, &
s_{F,\rS} &= s_F + {1 \ov 2} \epsdef \fm_F \,.
\eea
and $s_{i,\rN/\rS} = q^i_F s_{F,\rN/\rS}$ inspired by the saddles \eq{sigma saddle} and \eq{tm saddles}. As before, the sum of $\fm$ is taken over the GNO quantized magnetic fluxes, while the product over $\alpha$ is taken over all positive roots.
$d_\text{grav}$, defined by
\be
d_\text{grav} = -\dim(\bfG)-\sum_i (r_i-1) \dim {\fR_i} \,,
\ee
coincides with the complex dimension of the target space when the gauge theory flows to an NLSM in the IR. $N_*$ is an integer, whose determination we do not get into here, while $t_s \in \fc^*_\bC$ is the complexified FI parameter shifted by a multiple of $1/2$, which amounts to the shift of the theta-angle by a multiple of $\pi$:
\be
t_s \equiv t + {1 \ov 2} \sum_{\alpha > 0 } \alpha
\quad (\text{mod}~\fh^*_\bZ) \,.
\ee

Some explanation is due regarding the evaluation of equation \eq{omega S2}. $\tfM^\fm_\sing$ denotes the codimension-$\rk(\bfG)$ singularities of the integrand $\mathbf{I}_\fm$ on $\bC^{\rk(\bfG)}$. The singularities lie where $s \geq \rk(\bfG)$ hyperplanes
\be
H^{\cI,k}_\fm = \Big\{~\hsig~:~
{Q_\cI (\hsig_\rN) + q_F^\cI s_{F,\rN}} + {r_\cI \epsdef \ov 2}
= -k \epsdef ~\Big\} \,,
\ee
for integers $k$ with
\be
0 \leq k \leq Q_\cI (\fm)  + q^\cI_F \fm_F-r_\cI
\ee
intersect. The indices $\cI$ and the variables defining the hyperplane equations are defined in equations \eq{cI def} and \eq{charges etc}. For each such singular point $\hsig_*$, we define, as before, the set of associated charges
\be
\bfQ(\hsig_*)  = \{ Q_{\cI_1}, \cdots, Q_{\cI_s}\} \,.
\ee
Now the JK residues at the codimension-$\rk(\bfG)$ poles may be evaluated by the choice of a JK vector. An important difference between this sphere partition function and the torus partition function is that this choice matters---it must be chosen to take the value
\be\label{xiUV}
\xi_\eff^\UV = \xi + {1\ov 2 \pi} b_0 \log R\,, \quad
R \gg 1 \,,
\ee
where $b_0$ is defined in equation \eq{beta}. When the IR fixed point is conformal with $b_0 = 0$, the meaning of $\xi_\eff^\UV$ is clear. To explain equation \eq{xiUV} for $b_0 \neq 0$, we must remind ourselves that the choice of the JK vector, at the end of the day, is choosing an $\rk(\bfG)$-dimensional chamber $\mathfrak{C}_{\xi,b_0}$ among the chambers separated by cones of dimension $< \rk(\bfG)$ spanned by the charges $Q_\cI$. Equation \eq{xiUV} instructs that $\mathfrak{C}_{\xi,b_0}$ should be chosen such that
\be
\exists R_0 > 1 ~\text{such that} \qquad
\xi + {1\ov 2 \pi} b_0 \log R~ \in~ \mathfrak{C}_{\xi,b_0} \qquad
\forall~R > R_0 \,.
\ee

As in the torus partition function, the formula \eq{omega S2} is not well-defined in the presence of non-projective singularities. The usual prescription of dealing with such cases---deforming the theory by some twisted masses to resolve the singularities and taking the limit where the masses vanish---applies here as well. Meanwhile, it may be the case that $\xi_\eff^\UV \in i \fc^*$ lies squarely on a lower-dimensional cone spanned by the charges. In this case, one should slightly deform the JK vector, possibly to lie in $i\fh^* \setminus i\fc^*$, and evaluate the formula.

Let us now write out correlators $\vev{\s^n \evon{\rN}}_\epsdef$ for theory (i). The integration measure for the correlator is given by
\be
\mathbf{I}_\fm =  d \left({\hsig \ov \epsdef}\right) \cdot
\left(\hsig-{1 \ov 2}\epsdef \fm \right) ^n  \cdot
\begin{cases}
\prod_{p=0}^\fm
\left( {\hsig \ov \epsdef} -{\fm \ov 2} + p \right)^{-N_f} & \fm \geq 0 \\
1 & \fm = -1 \\
\prod_{p=1}^{-\fm-1}
\left( {\hsig \ov \epsdef} +{\fm \ov 2} + p \right)^{N_f} & \fm \geq -2
\end{cases}
\ee
for the magnetic flux $\fm$, which is now just an integer. We see the integration measure does not have any poles when $\fm <0$ and thus the sum over fluxes can be taken over non-negative integers. Meanwhile, $b_0 = N_f >0$. Since all the charges of the matter are given by $Q_\cI =1 > 0$, this means we need to sum over all the poles of the integrand $\mathbf{I}_\fm$. After introducing the variable $x= (\hsig/\epsdef -\fm/2)$, we arrive at the formula:
\bea
\vev{\s^n \evon{\rN}}_\epsdef &= \epsdef^{1-N_f}
\sum_{\fm = 0}^\infty {q^\fm \ov \epsdef^{N_f \fm}} \sum_{\ell = 0}^\fm
\Res{x = -\ell} {\epsdef^n x^n \ov \prod_{p=0}^\fm (x+p)^{N_f}} \\
&= - \epsdef^{1-N_f}
\sum_{\fm = 0}^\infty  {q^\fm \ov \epsdef^{N_f \fm}}
\Res{x = \infty} {\epsdef^n x^n \ov \prod_{p=0}^\fm (x+p)^{N_f}} \,.
\eea
We can explicitly evaluate the residues to obtain
\be
\vev{\s^n \evon{\rN}}_\epsdef =
\begin{cases}
0 & n \leq N_f -2 \\
1 & n = N_f -1 \\
0 & n = N_f \,.
\end{cases}
\ee
For larger $n$, we may use the following identity
\bea\label{CPNf def 0}
\vev{\s^{N_f} f(\s) \evon{\rN}}_\epsdef &=
-\epsdef^{1 -N_f}
\sum_{\fm} {q^\fm \ov \epsdef^{N_f}}
\Res{x = \infty} {\epsdef^{N_f} x^{N_f} f(\epsdef x)
\ov \prod_{p=0}^\fm (x+p)^{N_f}} \\ 
&= -q \epsdef^{1-N_f}
\sum_{\fm} {q^{(\fm-1)} \ov \epsdef^{N_f (\fm-1)}}
\Res{x = \infty} {f(\epsdef x -\epsdef) \ov \prod_{p=0}^{\fm-1} (x+p)^{N_f}} \\
&= q \, \vev{f(\s-\epsdef) \evon{\rN}}_\epsdef \,,
\eea
to compute the expectation values, where we shifted the variable $x \ra x-1$ in the second line of the equation. This is a non-associative deformation of the quantum cohomology ring \cite{WittenGW,Witten:1989ig,Intriligator:1991an} of $\mathbb{CP}^{N_f-1}$, as is discussed further in section \ref{ss:geometry}.

The correlators for theory (ii), the quintic GLSM can be similarly written, where we set ourselves in the geometric phase of the theory $\xi > 0$:
\be\label{quintic correlators}
\vev{\s^n \evon{\rN}}_\epsdef
= \epsdef^{n-3}
\sum_{\fm = 0}^\infty \Res{x = \infty}
{\prod_{j=0}^{5\fm} (-5 x -j) \ov
\prod_{p=0}^\fm (x+p)^5} x^n \,.
\ee

We now write down correlators for theory (iii). The formulae being quite long, we set $\epsdef = 1$, and use the packaged variables extensively. We define
\bea
\s_{a,\rN} &= \hsig_a - {\fm_a \ov 2}\,, &
\s_{a,\rS} &= \hsig_a + {\fm_a \ov 2}\,, \\
\S_{F,\rN} &= s_F - \left( {\fm_F -r_F \ov 2} \right) \,, &
\S_{F,\rS} &= s_F + \left( {\fm_F -r_F \ov 2} \right) \,, \\
\cS_{A,\rN} &= \cs_A - \left( {\widecheck{\fm}_A+r_A \ov 2} \right)\,, &
\cS_{A,\rS} &= \cs_A + \left( {\widecheck{\fm}_A+r_A \ov 2} \right)\,,
\eea
and the traces
\be
\S_{\rN/\rS} = \sum_a \s_{a,\rN/\rS} \,.
\ee
As in the case of the round sphere partition function, we also introduce the following differences:
\bea
\S^a_{b,\rN/\rS} &= \s_{a,\rN/\rS}-\s_{b,\rN/\rS}\,, &
\S^a_{F,\rN/\rS} &= \s_{a,\rN/\rS}-\S_{F,\rN/\rS}\,, \\
\S^a_{A,\rN/\rS} &= \s_{a,\rN/\rS}-\cS_{A,\rN/\rS}\,, &
\S^{F_1}_{F_2,\rN/\rS} &= \S_{F_1,\rN/\rS}-\S_{F_2,\rN/\rS}\,, \\
\S^F_{A,\rN/\rS} &= \S_{F,\rN/\rS}-\cS_{A,\rN/\rS}\,, &
\S^{A_1}_{A_2,\rN/\rS} &= \cS_{A_1,\rN/\rS}-\cS_{A_2,\rN/\rS}\,.
\eea

As explained before, the operators we concern ourselves with can be expressed as Weyl-invariant polynomials of the eigenvalues of the sigma fields. In the case of the $U(N)$ theory, these are none other than the symmetric polynomials of $N$ variables. The correlators then can be computed to give
\bea
\vev{ \cO_1 \evon{\rN}\cO_2 \evon{\rS}}_\epsdef
= {(-1)^{N_*} \ov N!}
\sum_{\fm \in \bZ^N} {e^{2\pi i t_s (\S_{\rS}-\S_\rN)}}
\sum_{\hsig_* \in \tfM^\fm_\sing}
\JKRes{\hsig = \hsig_*} [\bfQ(\hsig_*),\xi^\UV_\eff]
\, \mathbf{I}_\fm \,,
\eea
where the integration measure $\mathbf{I}_\fm$ is given by
\bea
\mathbf{I}_\fm=&
\bigwedge_{a=1}^N d \hsig_a
\cdot \cO_1(\S_{a,\rN}) \cdot \cO_2(\S_{a,\rS}) \\
&\cdot \prod_{a<b} \left( \S^a_{b,\rN} \S^a_{b,\rS}\right)
\prod_{a=1}^N \left(
\prod^{N_f}_{F=1} {\Ga(\S^a_{F,\rN}) \ov \Ga(\S^a_{F,\rS}+1)}
\prod^{N_a}_{A=1} {\Ga(-\S^a_{A,\rN}) \ov \Ga(-\S^a_{A,\rS}+1)}
\right) \,.
\eea

Taking the FI parameter $\xi$ to be positive, we find that the poles picked up by the contour integral are located at
\be
\S_{a,\rN} = S_{F_a,\rN}-n_{a,\rN} \,, \qquad
\S_{a,\rS} = S_{F_a,\rS}+n_{a,\rS}
\ee
for some $\vF \in C(N,N_f)$ for non-negative integers $n_{a,\rN/\rS}$. At the end of the day, the integral factorizes, much like the round sphere partition function, into the form
\be
\sum_{\vF \in C(N,N_f)}
\cZ_0^\vF \cZ_\rN^{\vF,\cO_1} \cZ_\rS^{\vF,\cO_2} \,.
\ee
The functions $\cZ_{\rN/\rS}^{\vF,\cO}$ are related to the vortex partition functions defined in equation \eq{VPF} by
\bea
\cZ^{\vF,\cO}_{\rN} &= Z_\text{v}^{\vF,\cO} (-\S_{F,\rN}; -\cS_{A,\rN}, (-1)^{N_f+N} q) \,, \\
\cZ^{\vF,\cO}_{\rS} &= Z_\text{v}^{\vF,\cO} (\S_{F,\rS}; \cS_{A,\rS}, (-1)^{N_f+N_a} q) \,.
\eea
We note that analogous results in higher dimensions have been obtained in Refs. \citen{Factorization3d, Factorization4d}.

The expectation values \eq{omega S2} reproduce the expectation values of operators in the $A$-twisted theory \cite{WittenTop} when $\epsdef$ is taken to zero:
\be\label{epsdef zero}
\lim_{\epsdef \ra 0}
\vev{\cO_1 \evon{\rN}\cO_2 \evon{\rS}}_{\epsdef}
= \vev{\cO_1 \cO_2}_A \,.
\ee
This is evident from the supersymmetric background \eq{omega S2 bg}. This computation can be viewed as the Coulomb-branch counterpart of the Higgs-branch computation of the $A$-twisted correlators carried out in Ref. \citen{MorrisonPlesser}. More discussion on these correlators from the geometric point of view is presented in section \ref{ss:geometry}.

The interpretation of the correlators on the $\Omega$-deformed sphere remains mysterious from the field theoretic point of view. The operation of composing operators at the poles becomes non-associative in the presence of $\epsdef$, which is evident, for example, in equation \eq{CPNf def 0} for correlators in the $\mathbb{CP}^{N_f-1}$ model. While one may wonder if this has to do with the fact that operators constructed out of the sigma fields preserve supersymmetry only when they are placed at the poles, no clear physical picture of the supersymmetric operators has been given yet.%
\footnote{Some hints on the nature of these correlators exist in the literature, for example, in the discussion about gravitational descendant invariants in chapters 26-30 of Ref. \citen{Hori:2003ic}.}
It would be desirable to gain an understanding of these correlators based on a solid framework comparable to that of the $A$-twisted theory.

\section{More backgrounds}
\label{s:other}

Before moving on to applications of localization computations on supersymmetric backgrounds, let us give a brief summary of backgrounds and partition functions that we have not been able to review in detail.

We begin with the hemisphere partition function computed in Refs. \citen{SugishitaTerashima, HondaOkuda, HoriRomo}. Since the hemisphere has a boundary, additional data living at the boundary must be introduced in addition to the supersymmetric background specified in the bulk. The appropriate data turns out to be a $\bZ_2$-graded hermitian Chan-Paton vector space, and a certain polynomial function related to the superpotential of the theory \cite{Kapustin:2002bi, Brunner:2003dc, Hori:2004zd, Herbst:2008jq}. Upon localizing the supersymmetric gauge theories introduced in section \ref{s:basics}, the hemisphere partition function turns out to be a function of the FI parameters, the twisted masses, and the Chan-Paton data. In fact, the Chan-Paton data specifies a $B$-brane\cite{Ooguri:1996ck, Hori:2000ck} $\mathfrak{B}$, while the partition function itself is conjectured to compute the central charge of that brane. This central charge can be understood as an overlap between the canonical Ramond-Ramond (RR) ground state and the RR state corresponding to the brane:
\be
Z_{D^2}(\mathfrak{B}) =
{}_\text{RR}
\langle \, \mathfrak{B} \,|\, 0 \, \rangle_\text{RR} \,.
\ee
When the gauge theory flows to a Calabi-Yau manifold, the $B$-branes can be thought of as D-branes wrapping holomorphic cycles of the manifold in the large-volume limit. Upon choosing a suitable basis of branes, both the round sphere and $\Omega$-deformed sphere partition functions can be written as a weighted sum over a product of hemisphere partition functions \cite{HondaOkuda,HoriRomo}, i.e., the hemisphere partition functions can be thought of as building blocks for sphere partition functions. Hemisphere partition functions have also been used to confirm the role of the gamma class \cite{Libgober,Iritani1,Iritani2,KKP} in computing the central charge of $B$-branes.%
\footnote{See also Refs. \citen{KLY, Halverson:2013qca, Galkin:2014laa}.}

Meanwhile, the $\mathbb{RP}^2$ partition function \cite{KLY}, when the IR theory of the gauge theory is a sigma model into a Calabi-Yau manifold, can be interpreted as the central charge of orientifold planes in the large-volume limit:
\be
Z_{\mathbb{RP}^2}(\mathfrak{C}) =
{}_\text{RR}
\langle \, \mathfrak{C} \,|\, 0 \, \rangle_\text{RR} \,.
\ee
Since $\mathbb{RP}^2$ is an unoriented manifold, there is no sum over fluxes when computing the partition function. The fundamental group of $\mathbb{RP}^2$, however, is nontrivial---it is $\bZ_2$. Thus the $\mathbb{RP}^2$ partition function must be computed by summing over the $\bZ_2$ valued holonomies of the gauge fields. Upon choosing the appropriate weight between the distinct topological sectors, one can compute the crosscap amplitude, or the central charges of space-time filling orientifolds for Calabi-Yau manifolds in the large-volume limit. When the gauge theory has a $\bZ_2$-valued flavor symmetry, a holonomy with respect to such a symmetry may be turned on along the $\bZ_2$ element of the fundamental group of the $\mathbb{RP}^2$. When the theory flows to an NLSM in the IR, the flavor symmetry implies the existence of a $\bZ_2$ isometry of the target manifold. The partition function with the flavor holonomy activated then turns out to compute the central charge of lower dimensional orientifold planes wrapping submanifolds located along the fixed points of the corresponding $\bZ_2$ isometry. 

The $A$-twisted partition function and correlators of a gauge theory on a closed, orientable Riemann surface $\Sigma_g$ of genus $g \geq 1$ have been computed in Ref. \citen{Benini:2016hjo}.%
\footnote{The 2d localization formula of Ref. \citen{Benini:2016hjo} can be obtained by dimensionally reducing the partition function of a three-dimensional gauge theory on $\Sigma_g \times S^1$. The 3d computation of Ref. \citen{Benini:2016hjo} also appears in Ref. \citen{Closset:2016arn}.}
The genus-$g$ partition function of the $A$-twisted theory can be understood as a $g$-point function of the handle-operator\cite{Witten:1989ig} which has been computed for gauge theories, for example, in Ref. \citen{Nekrasov:2014xaa}. The localization computation correctly reproduces this result.

\section{Applications}
\label{s:applications}

Up to now, we have described various supersymmetric backgrounds that may be utilized to compute exact correlation functions of $\cN= (2,2)$ gauge theories. While the fact that we are able to compute expectation values of gauge theory observables exactly is satisfying in and of itself, it has further reaching physical and mathematical applications. While we mainly focus on applications of of supersymmetric localization to the study of 2d dualities (section \ref{ss:dualities}), and to quantum cohomology (section \ref{ss:geometry}), we have collected other important applications and point to the relevant literature in section \ref{ss:more applications}.

\subsection{Dualities}
\label{ss:dualities}

Duality refers to either the equivalence of two different Lagrangian theories under the map of their parameters, or the equivalence of their subsectors. In this section, we concern ourselves with infra-red dualities, which implies the equivalence of the IR fixed points, or even the IR effective theories of two distinct Lagrangian theories. The exact partition function or correlators of supersymmetric gauge theories can be used to confirm such dualities. While there are many dualities of two-dimensional $\cN=(2,2)$ gauge theories that have been proposed and studied \cite{HoriVafa, HoriTong, Hori:2011pd, BC, BPZ, GLF, Putrov:2015jpa, Gadde:2015wta}, we choose to focus on Hori-Vafa duality \cite{HoriVafa} and cluster dualities of quiver theories with unitary gauge group factors in this section. We also briefly touch upon dualities of theories with adjoint matter at the end of the subsection.

Hori-Vafa duality \cite{HoriVafa} refers to the equivalence between gauge theories with chiral matter and Landau-Ginzburg (orbifold) models of twisted chiral fields. More precisely, it refers to the duality between a gauge theory, specified by the data given in section \ref{s:basics}, with a theory of twisted chiral fields with the following data:
\begin{itemlist}
  \item $\Sigma$, with bottom component $\sigma$, is a twisted chiral field valued in the Cartan subalgebra $\fh_\bC$ of the gauge algebra $\fg$ of the original theory. We use the notation $\tr_I \s$ to denote the projection of $\s$ to the element $I$ of $\fc_\bC \in \fh_\bC$.
  \item For each chiral field $\Phi^i$ in the original theory, there is a corresponding set of twisted chiral fields $Y^{i,\rho}$, with bottom component $y^{i,\rho}$, labeled by the weights $\rho$ of $\fR_i$. $Y^{i,\rho}$ are periodic, i.e., $Y^{i,\rho} \sim Y^{i,\rho} + 2 \pi i$.
  \item The twisted superpotential is given by
  \bea
  \hW(\s,y) = \ha \sum_I t_I \, \tr_I \s
  -{i \ov 4 \pi} \sum_a \s^a \sum_i \sum_{\rho \in \Lambda_{\fR_i}} \rho^a y^{i,\rho}
  - {i \ov 4 \pi } \sum_i \sum_{\rho \in \Lambda_{\fR_i}} e^{-y^{i,\rho}}
  \eea
  \item The Weyl group $\cW$ of $\bfG$ is a discrete global symmetry of the Landau-Ginzburg theory of $\Sigma$ and $Y$. When $\cW$ is non-trivial, it is gauged. 
\end{itemlist}
Here we have ignored possible twisted masses and $R$-charges of the chiral fields in the original theory, but they are straightforward to incorporate. Hori-Vafa duality is the statement that correlators of the original gauge theory is reproduced by the correlators of this Landau-Ginzburg theory with an additional insertion of the operator
\be
\prod_{\alpha > 0} |\alpha(\s)|^2 \,,
\ee
to the path integral, where $\alpha$ runs over the positive roots of $\fg$. Note that while the algebra $\fg$ is used in defining this Landau-Ginzburg theory, it is not a gauge symmetry of the theory.

Now the round sphere partition function of the Landau-Ginzburg theory can be computed to be compared with that of the gauge theory. The partition function of twisted chiral fields, which we schematically denote by $Y$ for the moment, localizes on saddles where the twisted chiral fields take constant values \cite{GL}:
\be
Z_{S^2} = \int dY d \overline{Y} e^{-4 \pi \hW (Y) + 4 \pi \overline{\hW (Y)}} \,,
\ee
where our conventions slightly differ from the original reference. We can then compute the sphere partition function of the Landau-Ginzburg theory that should match that of the gauge theory---it is given by
\bea
Z^{LG}_{S^2} ={1 \ov |\cW|} &\int d \s d \overline{\s}
\prod_{\alpha > 0} |\alpha(\s)|^2
e^{-2\pi t(\sigma)+2\pi \bar{t}(\bar{\s})} \\
&\cdot \prod_i \prod_{\rho \in \Lambda_{\fR_i}}
\left( \int dy_\rR \int_{-\pi}^{\pi} dy_\rI \,
e^{2i \rho(\s_\rR) y_\rR + 2 i \rho(\s_\rI) y_{\rI} }
e^{2i e^{-y_\rR} \sin y_\rI} \right)\,,
\eea
where the factor of $1/|\cW|$ in the front of the equation is due to the orbifolding action. We have introduced the subscripts $\rR$ and $\rI$ to denote the real and imaginary part of the variables involved. Now note that unless $\rho(\s_\rI)$ are half-integers, the $dy_\rI$ integrals vanish. Thus we find that the imaginary part of the sigma fields must be GNO quantized:
\be
\s = \hsig - {i \ov 2} \fm\,,\quad
\rho(\fm) \in \bZ \,.
\ee
The integral can now be written as a sum over GNO quantized fluxes:
\bea
Z^{LG}_{S^2}
= {1 \ov |\cW|} \sum_\fm  &\int d \s d \overline{\s}
\prod_{\alpha > 0} |\alpha(\s)|^2
e^{-4 \pi i \xi(\hsig) + i \theta(\fm)} \\
&\cdot \prod_i \prod_{\rho \in \Lambda_{\fR_i}}
\left( \int dy_\rR \int_{-\pi}^{\pi} dy_\rI \,
e^{2i \rho(\hsig) y_\rR - i \rho(\fm) y_{\rI} }
e^{2i e^{-y_\rR} \sin y_\rI} \right) \,.
\eea
The $y$ integrals can be carried out explicitly to reproduce the sphere partition \eq{ZS2} exactly. While Hori-Vafa duality was proven for abelian gauge theories in the original work, the non-abelian case was best described as a conjecture except in a limiting number of examples \cite{Bertram:2003qd}. The sphere partition function computation provides strong evidence for it being true for gauge theories in general.

\begin{figure}[b!]
\centerline{\includegraphics[width=10cm]{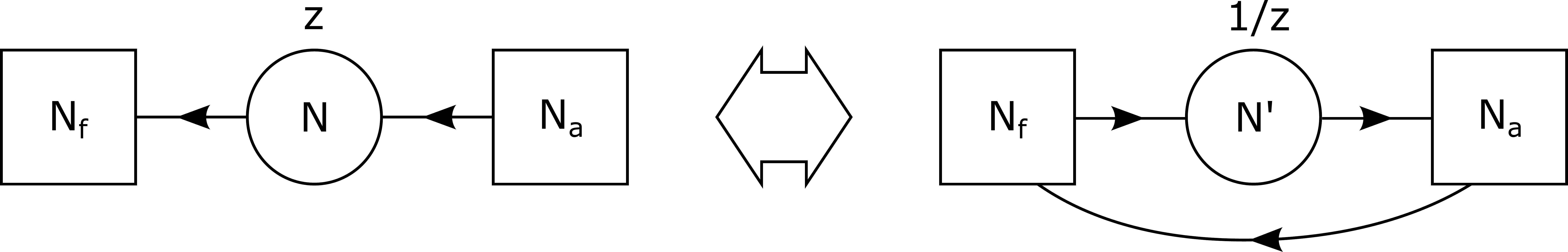}}
\caption{The quiver diagram of the dual theories (iii) (left) and (iv) (right). \label{f:UN}}
\end{figure}

Let us now discuss cluster duality \cite{BPZ}, elements of which have appeared in Refs. \citen{HoriTong,Hori:2011pd,BC,Hanany:1997vm}. Cluster duality \cite{BPZ}, some crucial components of were also discovered in Ref. \citen{GLF}, is based on the Seiberg-like duality \cite{SeibergDuality} of $U(N)$ theories with $N_f$ fundamentals and $N_a$ anti-fundamentals, i.e., theory (iii) \cite{HoriTong,BC}. The claim is that theory (iii) is dual to theory (iv):
\begin{enumerate}[(i)]
\setcounter{enumi}{3}
\item $U(N_f -N)$ theory with $N_f$ antifundamental and $N_a$ fundamental matter.
  \begin{enumerate}[(a)]
   \item Gauge group: $U(N')$ with $N' = N_f - N$.
   \item Charged matter: $N_f$ antifundamental chiral fields $Q'_F$, labeled by $F$ and $N_a$ antifundamental chiral fields $\widecheck{Q}'_A$, labeld by $A$. The $U(1)_R$ charges are given by $1-r_F$ and $1-\widecheck{r}_A$.
   \item There is a single chiral meson $M$, that transforms as a fundamental in the $U(N_f)$ subgroup and as an antifundamental in the $U(N_a)$ subgroup of the flavor symmetry. When generic twisted masses and $R$-charges are assigned, the flavor symmetry group breaks up into $U(1)^{N_f +N_a -1}$ and the meson breaks up into $N_f \times N_a$ massive chiral fields.
   \item The superpotential is given by
   \be
   W' = W(M_{AF})+ \sum_{F,A} M_{AF} Q'_F \widecheck{Q}'_A \,,
   \ee
   where $W$ is the superpotential of theory (iii), with the gauge invariant mesons $\widecheck{Q}_A Q_F$ of theory (iii) replaced by the singlets $M_{AF}$.
   \item The twisted superpotential of the theory is given by
   \bea\label{hW dual}
   \hW' = \ha t' \tr \s + \ha \Bigg[&
   \left( t+{N' \ov 2} \right) \sum_F s_F + 
   {N' \ov 2} \sum_A \cs_A \\
   &+{1 \ov 2 \pi i} \delta_{N_f,N_a} \ln (1+z) 
   \left( - \sum_F s_F  +\sum_A \cs_A \right) \Bigg] \,,
   \eea
   for $t' = -t +N_a/2$, where it is useful to recall that the twisted masses lie within a background vector multiplet. We remind the reader that we always assume that $N_f \geq N_a$. We have ignored various contact terms that do not depend on the (dynamical/background) vector multiplets. Here we have defined a convenient parameter
   \be
   z = e^{i \pi (N_f - N)}e^{2 \pi i t}
   \ee
   to make the equation simpler.
  \end{enumerate}
\end{enumerate}
Much of the data regarding the duality can be succinctly captured by a quiver diagram. A quiver diagram is made up of circular and square nodes with inscribed positive integers and directed edges which connect a pair of nodes. The nodes encode the gauge and flavor symmetry group---the circular nodes with inscribed numbers $N_p$ stand for the $U(N_p)$ gauge group factors, while the squares stand for flavor subgroups. Meanwhile, each edge corresponds to bifundamental matter, that is in the fundamental representation of the group at the tail, and an antifundamental representation of the group at the head. To each gauge node, we also associate a complex number $z$ that encodes the FI parameter of the corresponding unitary gauge group factor. The type of gauge theories whose gauge/matter content can be encoded into such a diagram is called a {\it quiver gauge theory}. The duality between theories (iii) and (iv) can then be expressed by figure \ref{f:UN}.

The bracketed terms of equation \eq{hW dual}, which are dependent on the twisted masses, have a surprising effect when the background flavor vector fields are promoted to dynamical gauge fields, i.e., when a subgroup of the flavor group is promoted to a gauge group. This happens when one takes a quiver gauge theory with unitary gauge group factors and dualizes the theory with respect to a gauge node, which we denote $p$ for now. The bracketed terms in equation \eq{hW dual} amount to shifting the FI parameters of the neighboring nodes of node $p$ by a function of $e^{2\pi i t_p}$, where $t_p$ is the complex FI parameter of $U(N_p)$. Needless to say, from the rules we have learned from the Seiberg-like duality of the $U(N)$ theory, the dual quiver theory should have a different rank, and a different quiver. Quite surprisingly, the rules for mutating the quiver and modifying the various FI couplings by this duality map, explained in all their glory in Ref. \citen{BPZ}, has been studied in detail in mathematics---they are precisely the mutation rules studied in cluster algebras, originally formulated by Fomin and Zelevinsky \cite{FominZ1}.

The bracketed terms in equation \eq{hW dual} were found in Refs. \citen{BPZ,GLF} by examining the sphere partition function of theory (iii). Recall from section \ref{s:round S2} that the sphere partition function of theory (iii) can be expressed as
\be
\sum_{\vF \in C(N,N_f)} Z_0^\vF Z_+^\vF Z_-^\vF \,, \qquad
Z_+^\vF = Z_\text{v}^\vF(\S_{F+} ;\cS_{A+};e^{2\pi i t}) \,,
\ee
where the vortex partition function $Z^{\vF}_\text{v} = Z^{\vF,\cO=1}_\text{v}$ is defined via equation \eq{VPF}. Now we can also write the sphere partition function of theory (iv) but {\it without} the additional terms proportional to the twisted masses in equation \eq{hW dual} by
\be
\sum_{\vFc \in C(N,N_f-N)} {Z'}_0^\vFc {Z'}_+^\vFc {Z'}_-^\vFc \,, \quad
{Z'}_+^\vFc = Z_\text{v}^\vF({1 \ov 2} -\S_{F+} ;-{1 \ov 2} -\cS_{A+};(-1)^{N_a}e^{2\pi i t}) \,,
\ee
where the ordered tuple $\vFc$, as a set, is the complement of $\vF$ with respect to $[N_f]$:
\be
\{ F^c_{I'} \} = [N_f] \setminus \{ F_I \} \,.
\ee
While the perturbative pieces satisfy the relation
\be\label{Z0 map}
Z_0^\vF \equiv
e^{2 \pi i\left( \left(t + {N' \ov 2} \right) \sum_F \S_{F+}
+ { N' \ov 2} \sum_A \cS_{A+} \right)}
e^{-2 \pi i\left(\left(\bar{t} + {N' \ov 2}\right) \sum_F \S_{F-}
+ { N' \ov 2} \sum_A \cS_{A-} \right)}
{Z'}_0^\vFc
\ee
up to an overall common factor independent of the twisted masses involved, the vortex partition functions $Z_\text{v}^\vF$ and $Z_\text{v}^\vFc$ satisfy the relation: 
\be\label{VPF map}
Z_+^\vF = {Z'}_+^\vFc \times
\begin{cases}
1 & N_f \geq N_a +2 \\
e^{-z} & N_f = N_a +1 \\
(1+z)^{-\sum_F \S_{F+} + \sum_A \cS_{A+} + (N_f -N)} & N_f =N_a \,.
\end{cases}
\ee
The relative factors in \eq{Z0 map} and \eq{VPF map} are precisely accounted for by the bracketed terms of equation \eq{hW dual} in the dual of the $U(N)$ theory.

The cluster dualities are IR dualities in a stronger sense in that the effective theory of the intermediate IR regimes of the dual theories, as well as their fixed points, are equivalent. In particular, when a unitary quiver gauge theory flows to an NLSM at an intermediate IR scale, the duality manifests itself as an equivalence between distinct constructions of the same target space. For example, the duality of the $U(N)$ theory has an interpretation as the canonical isomorphism of the Grassmannian \cite{WittenGr,DonagiSharpe,Sharpe:2015vza}. When the theory is conformal, the duality rules indicate how the coordinates of the quantum-corrected K\"ahler moduli space of a Calabi-Yau manifold are mapped under such equivalences. Cluster dualities of quiver theories also have been approached from the point of view of the gauge/Yang-Baxter equation correspondence \cite{Yamazaki:2013nra} using the torus partition function in Refs. \citen{Yagi:2015lha,Yamazaki:2015voa}.

Dualities of $U(N)$ gauge theories with adjoint matter have also been explored using round sphere partition functions. Duality of $\cN=(2,2)^*$ theories, which are $\cN = (4,4)$ theories broken by a twisted mass, have been studied in Refs. \citen{BPZ,GLF}. Meanwhile, Kutasov-Schwimmer-like dualities \cite{Kutasov:1995np} of theories with adjoint matter can also be verified by similar methods \cite{GLF}. These dualities have a beautiful application in the context of the famed Alday-Gaiotto-Tachikawa (AGT) correspondence \cite{AGT}, as the sphere partition function of certain $\cN=(2,2)$ theories can be identified with correlation functions of certain conformal field theories on a Riemann surface. We briefly discuss this point in section \ref{ss:more applications}.

\subsection{Geometric applications}
\label{ss:geometry}

In may instances, the $\cN=(2,2)$ gauge theories of study flow to non-linear sigma models of a K\"ahler manifold, which we denote $X$ throughout this section, in the IR. The various partition functions and correlators compute geometric quantities of the target space $X$ of the IR theory. The most basic example of this is the torus partition function. As reviewed in section \ref{s:torus}, the Euler character of the target space geometry of the IR theory is encoded in the torus partition function.

As noted earlier, when the gauge theory flows to an NLSM of a Calabi-Yau manfold, the round sphere partition function computes the quantum K\"ahler potential of the twisted chiral conformal manifold of the theory \cite{JKLMR,GL,GHKSST}:
\be
Z_{S^2} = e^{-K(t_I,\bar{t}_I)}\,.
\ee
This has rather profound geometric implications. In particular, when the Calabi-Yau manifold happens to be complex-three-dimensional, all the genus-zero Gromov-Witten invariants \cite{Gromov,DSWW,WittenGW}, which ``count" pseudo-holomorphic curves of given degree, can be extracted from this partition function \cite{JKLMR}. Recall that the exponentiated FI parameters
\be
q_I = e^{2 \pi i t_I}
\ee
parametrize the K\"ahler moduli space of the target manifold $X$ of the IR theory. These coordinates are often called ``algebraic coordinates" of the K\"ahler moduli space \cite{MorrisonPlesser}. The coordinates $q_I$ are natural from the point of view of the gauge theory, as they are straightforwardly related to physical UV couplings, in which the various partition functions are readily expressed.

Meanwhile, there is a separate set of coordinates on this moduli space, denoted the ``flat coordinates" \cite{flat1,flat2,flat3}, that are more appropriate to extracting the Gromov-Witten invariants of the manifold. These flat coordinates $x^I$ are related to $q_I$ by the ``mirror map" of the form
\be\label{q to x map}
x^I = {\log q_I \ov 2 \pi i} +x_0^I + f^I(q)
\ee
where $f^I$ is a holomorphic function of the algebraic coordinates. The constants $x_0^I$ and the functions $f^I$ can be extracted from the fact that the quantum K\"ahler potential, or the sphere partition function of the theory, is given by the form \cite{JKLMR}%
\footnote{In order to arrive at the given formula, an appropriate frame must be chosen, or equivalently, a product of a holomorphic and antiholomorphic function of the $q_I$ coordinates must be multiplied to the sphere partition function:
\be
e^{-K} = F(q_I) \overline{F(q_I)} Z_{S^2} \,.
\ee
The choice of the appropriate function $F(q_I)$ requires the knowledge of the Euler character $\chi(X)$, which can be obtained by computing the torus partition function.}
\bea
e^{-K(x^I,\bar{x}^I)}
=&-{i \ov 6} \sum_{I,J,K} \kappa_{IJK}
(x^I -\bar{x}^I)(x^J-\bar{x}^J)(x^K -\bar{x}^K)
+{\zeta(3) \ov 4 \pi^3} \chi(X) \\
&+{2 i \ov (2\pi i)^3} \sum_\eta N_\eta
\left( \text{Li}_3 (e^{2\pi i x \cdot \eta}) + \text{Li}_3 (e^{-2\pi i \bar{x} \cdot \eta})\right) \\
&-{i \ov (2\pi i )^3} \sum_{\eta,I}
N_\eta
\left( \text{Li}_2 (e^{2\pi i x \cdot \eta}) + \text{Li}_2 (e^{-2\pi i \bar{x} \cdot \eta})\right)
\eta_I (x^I - \bar{x}^I)
\eea
where $\eta$ runs over the elements of $H_2(X,\bZ)$, $\chi(X)$ denotes the Euler character of $X$, and
\be
\text{Li}_k (z) = \sum_{n=1}^\infty {z^n \ov n^k} \,.
\ee
The numbers $N_\eta$ are the integral genus-zero Gromov-Witten invariants labeled by the homology class $\eta$. Thus, once the map \eq{q to x map} is established, it can be inverted to write the quantum K\"ahler potential in the flat coordinates, from which the Gromov-Witten invariants can be extracted. Some applications of the round sphere partition function in this context can be found in Refs. \citen{Bonelli:2013rja, Bonelli:2013mma,Nawata:2014nca, Gerhardus:2016iot}.

As noted in section \ref{s:omega S2}, the $A$-twisted correlation functions of operators in the twisted chiral ring can be computed by taking the $\epsdef \ra 0$ limit from the localization on the equivariant $A$-twisted sphere. The operators studied in section \ref{s:omega S2}, i.e., the gauge-invariant polynomials of sigma fields, can be identified as elements of the ``vertical" cohomology
\be\label{vertical}
\bigoplus_{n=0}^{\text{dim} X} H^{n,n} (X)
\ee
of the target manifold $X$ \cite{WittenGr}, which forms a subset of the $A$-twisted operators of the non-linear sigma model. The vector space of gauge-invariant polynomials of the sigma fields have a natural grading, which is the degree of the polynomials with respect to the elements of $\sigma$. This grading can be identified with the grading $n$ of the vertical cohomology of equation \eq{vertical}.

The $A$-twisted correlators satisfy quantum cohomology ring \cite{Vafa:1991uz} relations, which is a deformation of the classical cohomology ring. To be concrete, let us consider the case when $X=\mathbb{CP}^{N_f-1}$. The classical cohomology of the theory is a ring generated by the hyperplane class, represented by the sigma field $\sigma$. Now $\sigma$, being a $(1,1)$ form, must satisfy
\be
\sigma^{N_f} =0 \,,
\ee
$\mathbb{CP}^{N_f-1}$ being complex $(N_f-1)$ dimensional. Thus the cohomology ring of $\mathbb{CP}^{N_f-1}$ is given by $\bZ[\sigma]/(\sigma^{N_f})$. In the quantum theory, however, this ring is deformed to $\bZ[\sigma]/(\sigma^{N_f}-q)$ with $q = e^{2\pi i t}$ \cite{WittenGW,Witten:1989ig,Intriligator:1991an}. This is realized in the $A$-twisted correlation functions:
\be\label{CPNf Q cohomology}
\vev{\sigma^{N_f} \cdot f(\sigma)}_{A} =
q \, \vev{f(\sigma)}_{A} \,,
\ee
which can be obtained from equation \eq{CPNf def 0} by taking $\epsdef \ra 0$. Here, $f(\s)$ is an arbitrary polynomial of $\sigma$. The localization formulae for $A$-twisted correlation functions for Calabi-Yau GLSMs also reproduce classic results obtained by employing mirror symmetry \cite{Candelas:1990rm} or by direct counting of holomorphic curves \cite{MorrisonPlesser}. For example, for the quintic GLSM, equation \eq{quintic correlators} reproduces the famous result
\be
\vev{\sigma^k}_A =
\begin{cases}
{5 \ov 1 + 5^5 q} &\text{when $k=3$} \\
0 &\text{otherwise,}
\end{cases}
\ee
when $\epsdef$ is taken to vanish.

Meanwhile, the $\epsdef \ra 0$ limit of the formula \eq{omega S2} has been used to compute $A$-twisted correlators of Calabi-Yau NLSMs that have not been computed before. For example, new correlation functions of operators when $X$ is the Gulliksen-Neg\r{a}rd (GN) manifold \cite{Gulliksen}, which is a submanifold of $\mathbb{P}^7 \times \Gr (2,4)$, have been obtained this way \cite{CCP}. Let us explain this example in a little bit more detail. The GLSM for the GN manifold is a $U(1) \times U(2)$ theory \cite{JKLMR0}. Thus the sigma field can be written as $\s = \s_1 \oplus \s_2$ with $\s_1 \in \fu(1)$ and $\s_2 \in \fu(2)$, and there exist two algebraic K\"ahler coordinates obtained by exponentiating the FI parameters, which we denote $z$ and $w$. The vertical cohomology of $\mathbb{P}^7 \times \Gr (2,4)$ is generated by the elements
\be
\tr_1 \s \,, \quad
\tr_2 \s \,, \quad
\tr_2 \s^2 \,,
\ee
where the subscript on the traces label the algebra with respect to which the trace is being taken. Note that the existence of the inherently non-abelian operator $\tr_2 \s^2$ is linked to the fact that $\Gr (2,4)$ is not toric. While the vertical cohomology of $\mathbb{P}^7 \times \Gr (2,4)$ is bigger, the basis elements of the vertical cohomology of $X \subset \mathbb{P}^7 \times \Gr (2,4)$ are given by six elements:
\bea
1 &\in H^{0,0}(X) \,,
&
(\tr_1 \s)^3\evon{X}
&\in H^{3,3}(X) \,, \\
\tr_1 \s\evon{X} \,, ~ \tr_2 \s\evon{X}
&\in H^{1,1}(X) \,, &
(\tr_1 \s)^2\evon{X} \,, ~ (\tr_2 \s)^2\evon{X}
&\in H^{2,2}(X) \,.
\eea
Here the notation ``$\,\evon{X}\,$" has been used to denote that the given cohomology class has been pulled back to, or restricted to, $X$. Now since the operator $\tr_2 \s^2$ flows to a four-form in the IR theory, it must be that
\be
\tr_2 \s^2 \evon{X} = a (\tr_1 \s)^2\evon{X} + b (\tr_2 \s)^2\evon{X}
\ee
for some coefficients $a$ and $b$, which are dependent on the two algebraic coordinates $z$ and $w$ on the K\"ahler moduli space of $X$. These coefficients can be computed exactly by solving the linear equations
\bea
a \, \vev{\tr_1 \s\evon{X} (\tr_1 \s)^2\evon{X}}_A
+ b\, \vev{\tr_1 \s\evon{X}  (\tr_2 \s)^2\evon{X}}_A &= 
\vev{\tr_1 \s\evon{X} \tr_2 \s^2 \evon{X}}_A \,, \\
a\,\vev{\tr_2 \s\evon{X} (\tr_1 \s)^2\evon{X}}_A
+ b\, \vev{\tr_2 \s\evon{X}  (\tr_2 \s)^2\evon{X}}_A &= 
\vev{\tr_2 \s\evon{X} \tr_2 \s^2 \evon{X}}_A \,.
\eea
All the correlations functions listed in this equation have been computed by taking the $\epsdef \ra 0$ limit of the correlators on the equivariant $A$-twisted sphere in Ref. \citen{CCP}, which thus leads to the values of $a$ and $b$. To the author's knowledge, this result has not been obtained before Ref. \citen{CCP}, as the computation of correlators involving non-abelian operators have only been performed for a limiting number of examples before the recent advances in 2d $\cN=(2,2)$ localization techniques.

Turning $\epsdef$ on has an interesting effect. As discussed earlier, the most conspicuous is that the quantum cohomology ring undergoes a non-associative deformation---the existence of $\epsdef$ renders the action of composition of fields, while still commutative, non-associative. A simple example is that the quantum cohomology relation of equation \eq{CPNf Q cohomology} is deformed into
\be
\vev{\s^{N_f} f(\s) \evon{\rN}}_\epsdef
= q \, \vev{f(\s-\epsdef) \evon{\rN}}_\epsdef \,,
\ee
as derived in section \ref{s:omega S2}. The correlators for Calabi-Yau GLSMs also become more interesting once $\epsdef$ is turned on. The correlators of the sigma fields for the quintic GLSM can be evaluated using equation \eq{quintic correlators}:
\bea
\vev{\s^k \evon{\rN}} &= 0 \quad ( k=0,1,2)\,, &
\vev{\s^3 \evon{\rN}} &= {5 \ov 1 + 5^5 q} \,, \\
\vev{\s^4 \evon{\rN}} &= \epsdef {2 \cdot 5^6 q \ov (1 + 5^5 q)^2} \,, &
\vev{\s^5 \evon{\rN}} &= \epsdef^2 {5^5 q (-17 + 13 \cdot 5^5 q) \ov (1 + 5^5 q)^3} \,, \\
&\vdots
\eea
The meaning of these correlators are not entirely clear from the geometric point of view, although it seems sensible to conjecture that they are computing equivariant classes of the moduli space of holomorphic maps from a two-punctured sphere to $X$. This moduli space has a natural $\bC^*$ action, and $\epsdef$ may be identified with the equivariant parameter with respect to this action \cite{GiventalQuintic}. More discussions along these lines can be found in Ref. \citen{Ueda:2016wfa}.

\subsection{More applications}
\label{ss:more applications}

We conclude with listing and providing references for some important topics we did not cover in the previous subsections.

One place that $\cN=(2,2)$ gauge theories appear is as worldvolume theories of surface operators in four-dimensional theories with $\cN=2$ supersymmetry \cite{Gukov:2006jk}.%
\footnote{A related context in which 2d $\cN=(2,2)$ gauge theories appear in the study 4d $\cN=2$ gauge theories can be found in Refs. \citen{Hanany:2003hp, Auzzi:2003fs, Dorey:2011pa, Chen:2011sj}. There, the 2d gauge theories are identified as effective theories of vortex strings in the Higgs branch of the 4d theories.}
When the $\cN=2$ theory is a gauge theory with gauge group $\cG$, an interesting class of surface operators can be described by a gauge theory whose flavor current is coupled to the four-dimensional dynamical gauge fields \cite{Gaiotto:2009fs,DGL,GGS}. The localization techniques discussed in this review have been used to compute the partition function of these coupled 2d-4d systems on various backgrounds. The supersymmetric index of the $\cN=2$ theories in the presence of surface defects have been computed in Refs. \citen{GG,Alday:2013kda,Bullimore:2014nla}. Meanwhile, $S^4$ partition functions of $\cN=2$ theories with surface operators wrapped around an $S^2 \subset S^4$ have been computed in Refs.  \citen{GLF,Lamy-Poirier:2014sea}. By the AGT correspondence \cite{AGT,AGTSurface}, these partition functions have an interpretation as correlation functions of certain 2d conformal field theories on Riemann surfaces. The various two-dimensional duality relations presented in section \ref{ss:dualities} can be interpreted as symmetries of these correlation functions \cite{GLF}.

The gauge-Bethe correspondence\cite{NS1,NS2} is the correspondence between physical observables of certain 2d $\cN=(2,2)$ gauge theories and integrable systems. The round sphere partition function \cite{Bonelli:2015kpa} and the $A$-twisted correlators  \cite{Chung:2016lrm} of these gauge theories have been studied and interpreted in this context.

Supersymmetric partition functions have also been used to study gauge theories that flow to NLSMs of unconventional geometries in the IR. Localization computations have been carried out in Refs. \citen{Nian:2014fma,Benini:2015isa} for gauge theories with semi-chiral multiplets, which flow to geometries with torsion in the IR \cite{Crichigno:2015pma}. The spectrum of string states for ALE and ALF spaces have been studied  using the torus partition function of gauge theories with both chiral and twisted chiral matter in Ref. \citen{Harvey:2014nha}. The Gromov-Witten invariants of non-commutative resolutions of singular spaces have been computed in Ref. \citen{Sharpe:2012ji} using the round sphere partition function.

\section*{Acknowledgments}
I thank Francesco Benini, Cyril Closset, Stefano Cremonesi, Jaewon Song and Peng Zhao for collaborating on work presented in this review---most of what I know about the subject has been gained through the experience of working with them. I should also thank Allan Adams, Marcos Crichigno, Tudor Dimofte, Ethan Dyer, Abhijit Gadde, Davide Gaiotto, Sergei Gukov, Kentaro Hori, Bei Jia, Peter Koroteev, Vijay Kumar, Josh Lapan, Jaehoon Lee, Sungjay Lee, Bruno Le Floch, Noppadol Mekareeya, Dave Morrison, Nikita Nekrasov, Wolfger Peelaers, Martin Ro\v{c}ek, Mauricio Romo, Eric Sharpe and Yuji Tachikawa for educating discussions on related topics over the years. I would like to thank Francesco Benini, Noppadol Mekareeya, Dave Morrison and Jaewon Song again for helpful comments on the draft, and especially Cyril Closset and Stefano Cremonesi for being kind and patient with me while I have riddled them with questions throughout the course of this work. I
also thank the Korea Institute for Advanced Study for hospitality while this work was being carried out. This work is supported by DOE grant DOE-SC0010008.

\bibliographystyle{ws-ijmpa}
\bibliography{2DReview}

\end{document}